\documentclass[aps,pra,floatfix,twocolumn,letterpaper]{revtex4}

\pdfoutput=1

\usepackage{graphicx}
\usepackage{amsmath}
\usepackage{hyperref}
\usepackage{latexsym}
\usepackage{color}
\usepackage{amssymb}
\usepackage{amsmath}
\usepackage{float}
\usepackage{times}


\begin{document}
\title{Near Threshold Effects on Recombination and Vibrational Relaxation in Efimov Systems}

\author{D. Shu, I. Simbotin, R. C\^ot\'e}
\affiliation{Department of Physics, University of Connecticut, Storrs, CT 06268, USA}

\begin{abstract}
We investigate the energy dependence of inelastic processes in systems
which possess Efimov states. We consider the three-body recombination
rate $K_3$ where three free atoms interact to produce an
atom--dimer pair, and the relaxation rate $K_{\rm rel}$
where an atom quenches a weakly bound state of a dimer
near an Efimov resonance to more deeply bound levels.  Using a model
capturing the key features of the Efimov problem, we
  identify new energy regimes for $K_3$, namely the NTR (Near
  Threshold Resonance) regime behavior $E^{-2}$ for negative
  scattering lengths and the NTS (Near Threshold Suppression) regime
  behavior $E^2$ for positive scattering lengths.  We also confirm a
  previously found oscillatory behavior of $K_3$ at higher energy
  $E$. Finally, we find that $K_{\rm rel}$ behaves as $E^{-1}$ in the
  NTR regime.
\end{abstract}
\maketitle

\section{Introduction}
Three-body problems have been studied in a variety of
  context, such as three-body Coulomb systems
  \cite{Kais1,Kais2,Kais3}, and nuclear three-body systems
  \cite{nuc1,nuc2,nuc3}. Efimov predicted that a system with three
particles may have a large number of trimer states even when the dimer
potential does not posses any bound states
\cite{Efimov_original,Efimov_original2,Physrep428_Braaten}.  The
existence of the Efimov trimer states requires the two-body scattering
length $a$ to be much larger than the characteristic range of the
two-body interaction $R_0$. Ultracold gases are ideal candidates for
studying Efimov physics since the scattering length $a$ can be tuned
using Feshbach resonances. When $a \to -\infty$, an Efimov state near
the three-body threshold will give a resonant enhancement for the
recombination rate $K_3$. This enhancement has been experimentally
observed as atom loss for a variety of systems
\cite{Nature440_315,PRL103_JRwill,Science326_Pollack,
  PRL105_Gross,PRL107_Berninger,PRL108_Wild,PRL111_Roy,
  PRL101_Ottenstein,PRL102_Huckans,PRL105_Nakajima,
  PRL103_Barontini,PRL111_Bloom,PRL112_Pires,PRL113_Tung,PRL112_Huang}.
When $a \to +\infty$, a similar enhancement has been observed for the
vibrational relaxation rate $K_{\rm rel}$ for collisions between atoms
and loosely bound dimers
\cite{PRL105_Nakajima,PRL105_Lompe,Natphys5_227,PRL111_Bloom,PRL102_Huckans}.
Resonant peaks for atom loss rates have also been observed for pure
ultracold atom gases with $a > 0$
\cite{Natphys5_586,PRL103_Gross,Science326_Pollack}, and an avalanche
mechanism has been proposed as an explanation in terms of
molecule-atom threshold resonances
\cite{Science326_Pollack,Natphys5_586}, though other experiments have
been conducted pointing to different conclusion
\cite{PRA90_Zenesini,PRA90_Hu}.  For $a \to +\infty$, interference
minima as a signature of Efimov states in three atom loss rates have
also been observed
\cite{PRL103_Gross,Science326_Pollack,Natphys5_586,PRL105_Gross,PRA88_Dyke}.

Efimov physics has been studied mostly in the zero energy
  limit, e.g., the recombination and relaxation rates near zero
  temperature, and only recently the energy dependence of these
  quantities has received attention
  \cite{NJP_Wang,PRL104_Wang,prx-2016-salomon,prl-2013-salomon,pra-2013-zinner,pra-2015-grimm}.
In this paper, we investigate the energy dependence of the three body recombination rate $K_3$ and relaxation 
rate $K_{\mathrm{rel}}$, paying special attention to the threshold behavior of $K_3(E)$ for $a<0$ 
and $K_{\mathrm{rel}}(E)$ for $a>0$ when an Efimov state is near the threshold. We also explore the behavior
of $K_3(E)$ for $a>0$ for specific values of $a$ leading to large suppression effects.

\section{The model}

In this work, we study the Efimov effect for the case of identical
bosons, BBB, where B denotes a neutral bosonic atom in its ground
state.  Our findings are however also applicable to other systems
(such as mixtures of the type BBX) which have a similar attractive
Efimov potential.  In addition, we only consider the case of total
angular momentum $J=0$, which is sufficient at low energy where
contributions of higher $J$ values are strongly suppressed.

Although the
long range Efimov states seem counterintuitive, their appearance can
be explained by the attractive $1/R^2$ behavior of the lowest
adiabatic hyperspherical potential for $R_0 \ll R \ll |a|$, where $R$
is the hyperradius, and $a$ and $R_0$ are the two-body scattering
length and characteristic interaction range, respectively.  The
attractive Efimov potential takes the form \cite{long_review}
\begin{equation}
   V_{\rm Ef}(R) = -\frac{s_0^2+1/4}{R^2}\;,
   \label{eq:V_Ef.}
\end{equation}
where $s_0= 1.00624$ is a universal constant. Following
\cite{long_review}, the appropriate reduced mass is implicitly included
in $V_{\rm Ef}(R)$; the same is true throughout the
article for all potentials.  According to Efimov
\cite{Efimov_original}, when $a \to \pm \infty$, the number of
three-body Efimov states is proportional to $\ln (|a|/R_0)$, and is
independent of the sign of $a$. However, the lowest adiabatic
hyperspherical potential depends on the sign of $a$. Specifically,
when $a<0$, the Efimov potential correlates in the asymptotic region
$R \gg |a|$ to the repulsive potential
\begin{equation}
   V_{\mathrm{asy}}(R) = \frac{\ell_{\mathrm{eff}}(\ell_{\mathrm{eff}}+1)}{R^2} \;,
\end{equation} 
with $\ell_{\mathrm{eff}}=\frac{3}{2}$, which is the lowest three-body
continuum channel.  Again, $V_{\mathrm{asy}}$ contains the reduced
mass.  In this case, the three-body recombination rate $K_3$ exhibits
a resonant enhancement when an Efimov state is  near the
three-body breakup threshold. When $a>0$, the effect of the Efimov
states on the three-body loss rates is quite different, which is due
to the fact that the Efimov potential correlates to the weakly bound
dimer channel.  Following Esry et al.~\cite{long_review}, we employ a
simplified model which captures the essential physics of the Efimov
effect.

\subsection{Single-channel model for $a<0$}
\label{theory_1_channel}
\begin{figure}
\includegraphics[width=1.0\columnwidth,clip]{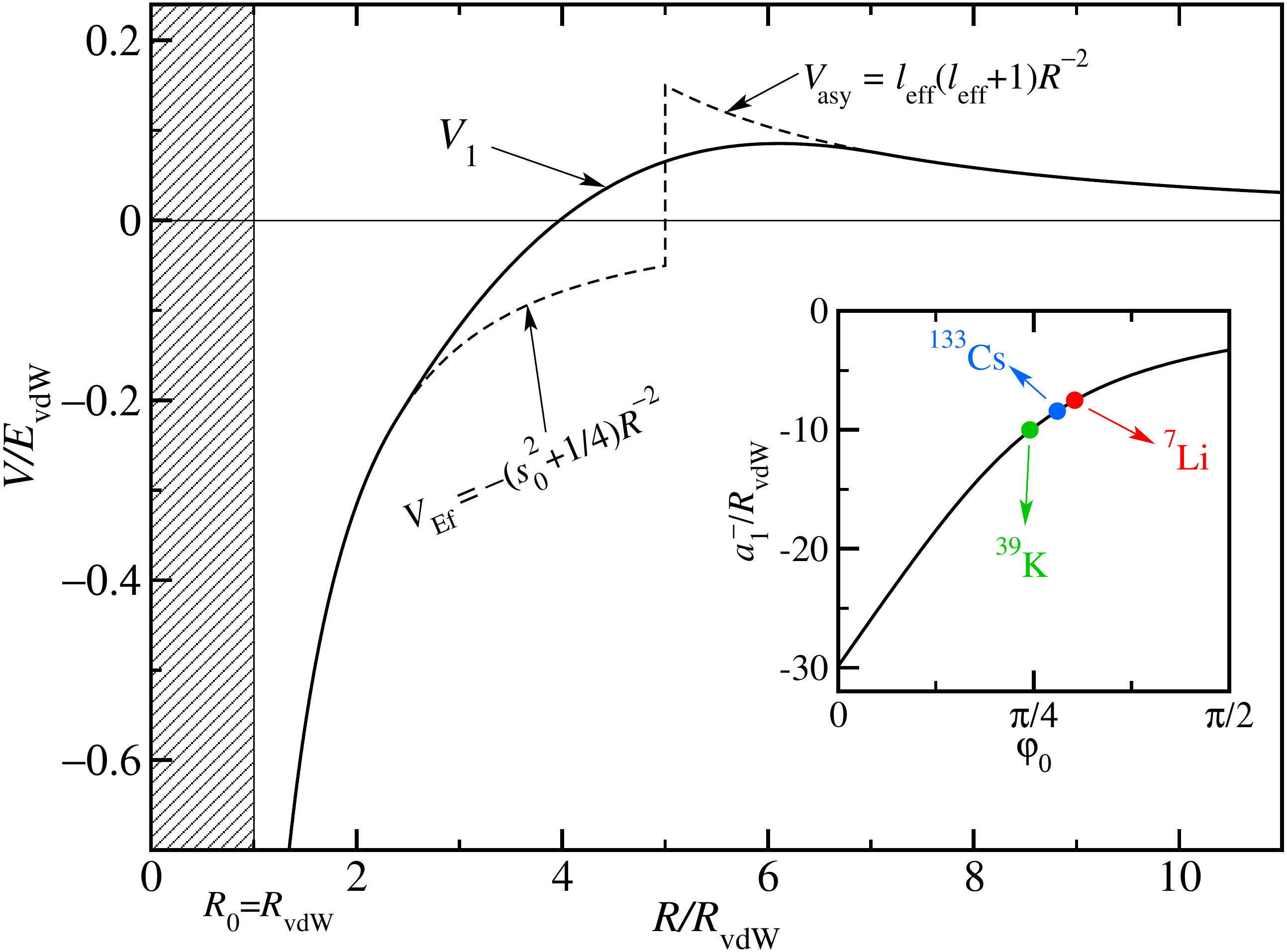}
\caption{\label{nega_a_potential} Lowest hyperspherical
  potential for $a=-5R_{\mathrm{vdW}}$. Solid line: smooth
  approximation for the realistic hyperspherical potential $V_1(R)$ in
  Eq.~(\ref{Radial_SE}). The shaded area represents the short range
  region $0< R < R_0$, taken into account via the
  phase $\varphi_0$ appearing as a parameter for the initial condition at
  $R_0=R_{\mathrm{vdW}}$ for Eq.~(\ref{Radial_SE}). Dotted line:
  approximate potential (discontinuous at $R=|a|$) used
  in Sec.~\ref{easy_case}. The inset shows the $\varphi_0$ dependence
  of the scattering length $a_1^-$ corresponding to the appearance of
  the first Efimov state. Also shown are three experimental results
  for Cs, K and Li \cite{long_review}  illustrating how 
  $\varphi_0$ can be adjusted to match experiments.}
\end{figure}

For $a<0$, the entrance channel is the lowest three-body
  continuum channel, and all other channels corresponding to the
  three-body continuum can be ignored since their hyperspherical
  adiabatic potentials are entirely repulsive; their contribution will
  be highly suppressed at low energy .  We assume that the dimer has
  deeply bound states, which is typical for atom--atom potentials, and
  we shall use a single channel model to analyze the three-body
  recombination.  The recombination channels corresponding to the
  deeply bound dimer states are taken into account \emph{a
    posteriori}, as explained in this section.  The fact that the
  coupling of the entrance channel to the recombination channels takes
  place at short range ($R<R_0$) allows for a  single-channel model
of the multi-channel problem that reproduces qualitatively the energy
dependence of $K_3$ \cite{NJP_Wang}. To extract $K_3(E)$ from the
single-channel results, we use an approach based on the Jost function,
as opposed to the wave function as used by \cite{NJP_Wang}. Namely, we
obtain the single-channel regular solution $\phi_k(R)$ by numerically
solving

\begin{equation}\label{Radial_SE}
\phi''_k(R) = [V_1(R)-k^2]\phi_k(R),
\end{equation}
  where $V_1$ is the potential curve shown in
  Fig.~\ref{nega_a_potential}. Note that for $R_0\ll R \ll |a|$,
  $V_1(R) \approx V_{\mathrm{Ef}}(R)$ and for $R \gg |a|$,
  $V_1(R)\approx V_{\mathrm{asy}}(R)$. Equation (\ref{Radial_SE}) is
  supplemented with the initial condition $\phi_k(R_0) = \sin
  \varphi_0$ and $\phi'_k(R_0) = \cos \varphi_0$, where the phase
  $\varphi_0$ accounts for the contribution of the short range
  region. As shown in the inset in Fig.~\ref{nega_a_potential}, the
  parameter $\varphi_0$ can be adjusted to obtain agreement with the
  experimental results for the value of $a_1^-$ of the two-body
  scattering length corresponding the appearance of first Efimov
  state. However, in the remainder of our paper, we employ the simple
  choice $\varphi_0=0$ which corresponds to  a hard wall at
  $R=R_0=R_{\mathrm{vdW}}$. The two-body van der Waals length
 $R_{\mathrm{vdW}}$ is the characteristic length scale for the short range
  region, with
  $R_{\mathrm{vdW}}=(2\mu_2 C_6/\hbar^2)^{1/4}$, where $\mu_2$ is the
  two-body reduced mass and $C_6$ the dispersion coefficient for the
  van der Waals interaction ($-C_6/R^6$) between two neutral ground
  state atoms. $R_{\mathrm{vdW}}$ and the corresponding van der Waals
  energy $E_{\mathrm{vdW}}=\hbar^2/2\mu_2 R_{\mathrm{vdW}}^2$ are used
  as units in Fig.~\ref{nega_a_potential}.

We extract the Jost function
\begin{equation}
   F(k) = A(k) - iB(k) 
\end{equation}
from the asymptotic ($R\to\infty$) behavior $k^{\ell_{\mathrm{eff}}+1}\phi_k(R) \sim
A(k)j_{\ell_{\mathrm{eff}}}(kR) + B(k)n_{\ell_{\mathrm{eff}}}(kR)$,
where $j_{\ell_{\mathrm{eff}}}$ and $n_{\ell_{\mathrm{eff}}}$ are the
Riccati--Bessel spherical functions.  Note that  $V(R)$ and $k$
are real, and thus $\phi_k$, $A(k)$ and $B(k)$ are also real. The physical
solution $\psi_k$, which behaves asymptotically as $\psi_k(R) \sim
\frac{i}{2}[h_{\ell_{\mathrm{eff}}}^{-}(kR) - \frac{F^*}{F}
  h_{\ell_{\mathrm{eff}}}^{+}(kR)]$, where $h_\ell^\pm \equiv n_\ell
\pm i j_\ell$, can be expressed as \cite{Taylor}
\begin{equation}\label{psi_F_phi}
\psi_k = \frac{k^{\ell_{\mathrm{eff}}+1}}{F(k)}\phi_k.
\end{equation}
For very low $k$, the regular solution is independent of $k$ at short range, 
$\phi_k(R) \approx \phi_0(R)$. Thus, at short range, we have 
\begin{equation}\label{short_range_phipsi}
\psi_k(R)|_{\mathrm{short\ range}} \approx \frac{k^{\ell_{\mathrm{eff}}+1}}{F(k)}\phi_0(R) , 
\end{equation}
Returning to the multi-channel problem, and noting that the coupling
of the entrance channel to the deeply bound states of the dimer takes
place at short range, the entrance channel component $\psi_{1,k}$ of
the full wave function can be approximated by the single-channel
solution (\ref{psi_F_phi}), i.e.,
\begin{equation}
   \psi_{1,k} \approx \psi_k = \frac{k^{\ell_{\mathrm{eff}}+1}}{F(k)}\phi_k   
   \label{eq:single-channel-psi_1k}
\end{equation}
Since the couplings are restricted to short range, the single channel
result in Eq.~(\ref{short_range_phipsi}) can be used to obtain the $k$
dependence of the full solution of the coupled-channel problem
\cite{Jost_paper}. Indeed, the $k$-dependence in
Eq.~(\ref{short_range_phipsi}) will be imprinted via the couplings to
all other components of the wave function. We emphasize that, although
the entrance-channel component has the $k$-dependence in
Eq.~(\ref{short_range_phipsi}) only at short range, the other
components obey this $k$-dependence for all $R$,
\begin{equation}\label{psi_nk}
   \psi_{n,k} \sim \frac{k^{\ell_{\mathrm{eff}}+1}}{F(k)} g_n(k_n,R) ,
\end{equation}
where $g_n(k_n,R)$ is the radial wave function for channel $n$ with
momentum $k_n$.  Recalling that only outgoing waves are allowed in the
dimer channels $n$, the corresponding asymptotic ($R \to \infty$)
behavior of $\psi_{n,k}$ is
\begin{equation}\label{psi_nk_asy}
   \psi_{n,k}(R) \sim \sqrt{\frac{k}{k_n}}S_{n,1}(k) e^{+ik_n R}.
\end{equation}
Together with Eq.~(\ref{psi_nk}) and substituting $2\ell_{\rm
  eff}+1=4$, it leads to the $k$-dependence of the S-matrix element,
\begin{equation}\label{Smat_F}
|S_{n,1}(k)|^2 \sim \frac{k^4}{|F(k)|^2}.
\end{equation}
Finally, the total three-body recombination rate \cite{PRL94_Incao},
$K_3 \sim \frac{1}{k^4}\sum_{n \neq 1} |S_{n,1}|^2$, reads
\begin{equation}\label{K3_F}
   K_3 \sim  \frac{1}{|F(k)|^2} = \frac{1}{A^2(k)+B^2(k)}.
\end{equation}
Thus, the $k$-dependence of the rate $K_3$ is dictated by the
$k$-dependence of the Jost function.

\subsection{Single-channel model for relaxation ($a>0$)}
\label{theory_1_channel_relax}

\begin{figure}[b]
\includegraphics[width=1.0\columnwidth,clip]{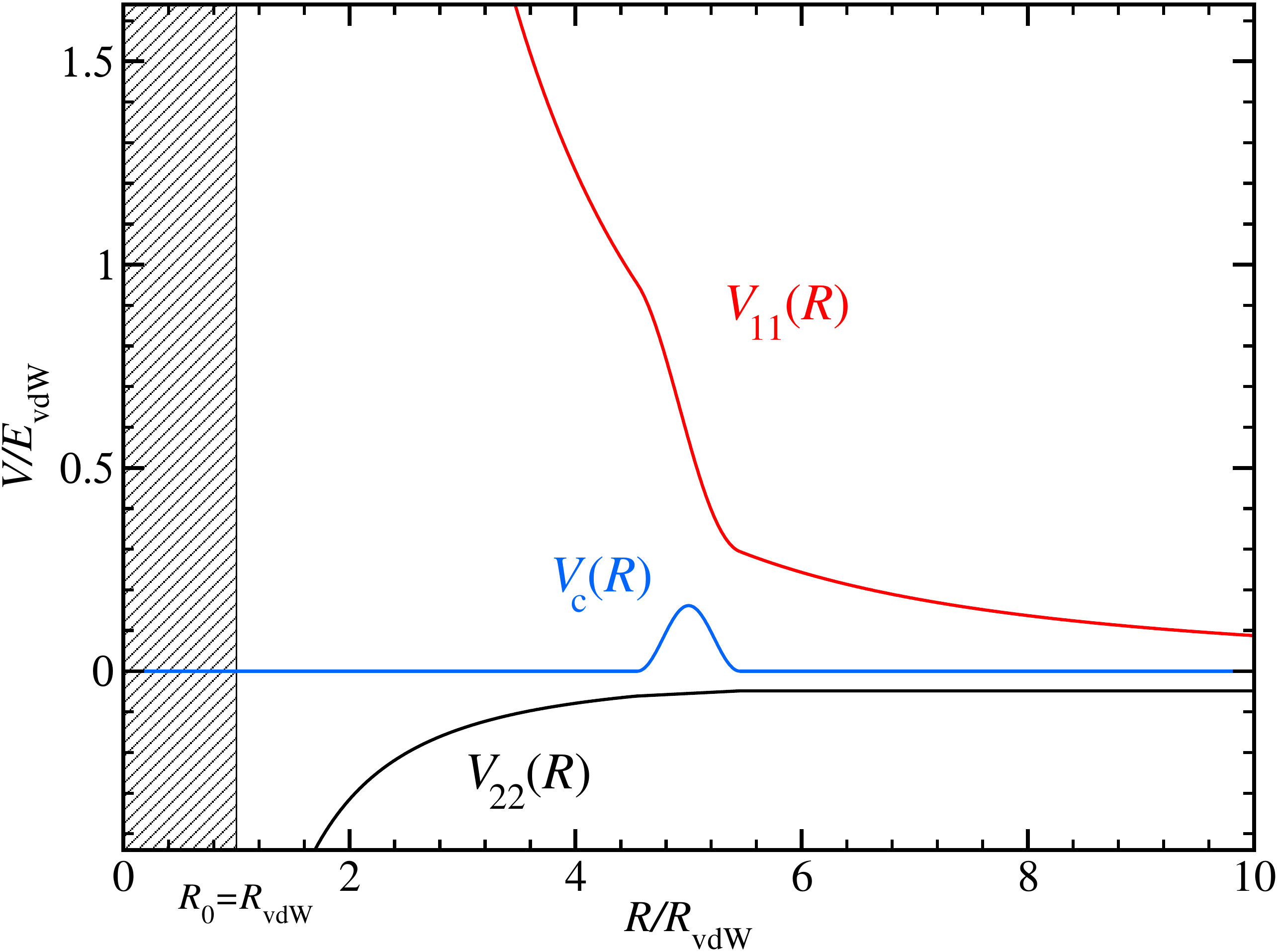}
\caption{\label{posi_a_potential}Schematic representation
  of the potentials used in the two channel model, see
  Eqs.~(\ref{eq:2channel_eq},\ref{eq:Vc}). Red curve: hyperspherical
  potential for channel 1. Black curve: hypershperical potential for
  channel 2 corresponding to the shallow dimer state. Blue curve:
  coupling potential $V_c$ which is non-zero only near $R=a$. The
  shaded area represents the short range region.  }
\end{figure}

As shown in Fig.~\ref{posi_a_potential} for the case $a>0$, there
exists a loosely bound dimer state (channel 2, slightly below the
three-body breakup threshold) which correlates to the Efimov
potential, while the lowest three-body continuum channel is purely
repulsive (channel 1). In channel 2, the three-body system corresponds
to an extended weakly bound Feshbach molecule interacting with an
atom. Since decay into more deeply bound and compact dimers takes
place due to short-range couplings, a single-channel model similar to
the case $a<0$ described in Sec.~\ref{theory_1_channel} above is
warranted for the rate of the vibrational relaxation rate
  $K_{\rm rel}$, which will thus be expressed in terms of the single
  channel Jost function.

In this case, we have an atom--dimer scattering problem
  with relative angular momentum $\ell$, so that one simply replaces
$\ell_{\rm eff}$ by $\ell$ in Eqs.~(\ref{eq:single-channel-psi_1k}) and
(\ref{psi_nk}), leading to
\begin{equation}
   |S_{fi}(k)|^2 \sim \frac{k^{2\ell +1}}{|F(k)|^2}.
   \label{eq:S-relax}
\end{equation}
We are interested in $\ell=0$ (ultracold regime), so that the cross
section $\sigma=\frac{\pi}{k^2}\sum_{f\neq i}|S_{fi}(k)|^2$ for this
inelastic process gives the relaxation rate $K_{\rm rel}=v_{\rm
  rel}\sigma$ (with the relative velocity $v_{\rm rel}\propto k$) in
terms of the S-matrix as \cite{PRL94_Incao}
\begin{equation}
   K_{\mathrm{rel}}(k) \sim \frac{1}{k}\sum_{f \neq i}|S_{fi}|^2,
\end{equation}
Here, $i$ and $f$ correspond to the entrance (shallow) dimer state,
and final (deeply bound) dimer states respectively. The S-matrix is
given in term of the Jost function by Eq.~(\ref{eq:S-relax}) with
$\ell=0$, i.e.,
\begin{equation}
   |S_{fi}(k)|^2 \sim \frac{k}{|F(k)|^2}.
\end{equation}
Combining the two equations above, we find the $k$-dependence of the
relaxation rate,
\begin{equation}
 K_{\mathrm{rel}}(k) \sim \frac{1}{|F(k)|^2}=\frac{1}{A^2(k)+B^2(k)}.
 \label{eq:K_rel}
\end{equation}

\subsection{Two-channel model for $a>0$}
\label{theory_2_channel}

As discussed above for $a>0$, and as illustrated in
Fig.~\ref{posi_a_potential}, there exists a loosely bound dimer state
(channel 2) correlated to the Efimov potential, while the lowest
three-body continuum channel being purely repulsive (channel
1). Therefore, to investigate the three-body recombination for the
formation of Feshbach molecules, we adopt a model that only includes
these two channels:
\begin{equation}
\begin{array}{rcl}
\phi''_1 &=& [V_{11} - k_1^2]\phi_1 + [V_c\partial_R + \partial_R V_c]\phi_2 \\
\phi''_2 &=& [V_{22} - k_2^2]\phi_2 - [V_c\partial_R + \partial_R V_c]\phi_1.
\end{array}\label{eq:2channel_eq}
\end{equation}
According to \cite{long_review}, the coupling between channel 1 and
channel 2 is significant only near $R=a$. Thus we employ a simple form
for the coupling,
\begin{equation} \label{eq:Vc}
V_c(R) = \left\{
\begin{array}{ll}
    \frac{0.01}{a} \sin^4\left(\pi\frac{R-R_a}{R_b-R_a}\right), & R_a<R<R_b \\
    0, & \mbox{otherwise} 
\end{array}
\right.
\end{equation}
with $ R_a = 0.5a, R_b = 1.5a$.  Note that the scaling $V_c(R) \sim
a^{-1}$ follows the realistic coupling, and the value
  $0.01$ for the coefficient controlling the strength of the coupling
  in Eq.~(\ref{eq:Vc}), chosen arbitrarily, can be adjusted
  to match experimental data for a particular system
\cite{long_review}.  In the region $R_0 \ll R \ll a$ we have
\begin{equation}
   V_{11}(R) \approx \frac{s_1^2-1/4}{R^2}, \qquad
   V_{22}(R) \approx V_{\mathrm{Ef}} = -\frac{s_0^2+1/4}{R^2} \;,
\end{equation}
where $s_1 = 4.4653$ and $V_{\mathrm{Ef}}$ is the Efimov potential in
Eq.~(\ref{eq:V_Ef.}).  In the asymptotic region $R \gg |a|$,
\begin{equation}
   V_{11}(R) \approx \frac{\ell_{\mathrm{eff}}(\ell_{\mathrm{eff}}+1)}{R^2},\qquad
   V_{22}(R) \approx E_b,
\end{equation} 
where $E_b\propto -1/a^2$ is the binding energy of the shallow dimer.
In the region near $R=a$, $V_{11}(R)$ and $V_{22}(R)$ are connected smoothly 
between the inner and outer regions using high order polynomials;
namely, we ensure their continuity and the continuity of their first and
second derivatives.

\section{Results}
In this section, we discuss the $k$-dependence of rate
  coefficients for the various cases introduced in the previous
sections.

\subsection{Single channel results for $a<0$} 
\label{easy_case}

\begin{figure}
\includegraphics[width=1\columnwidth,clip]{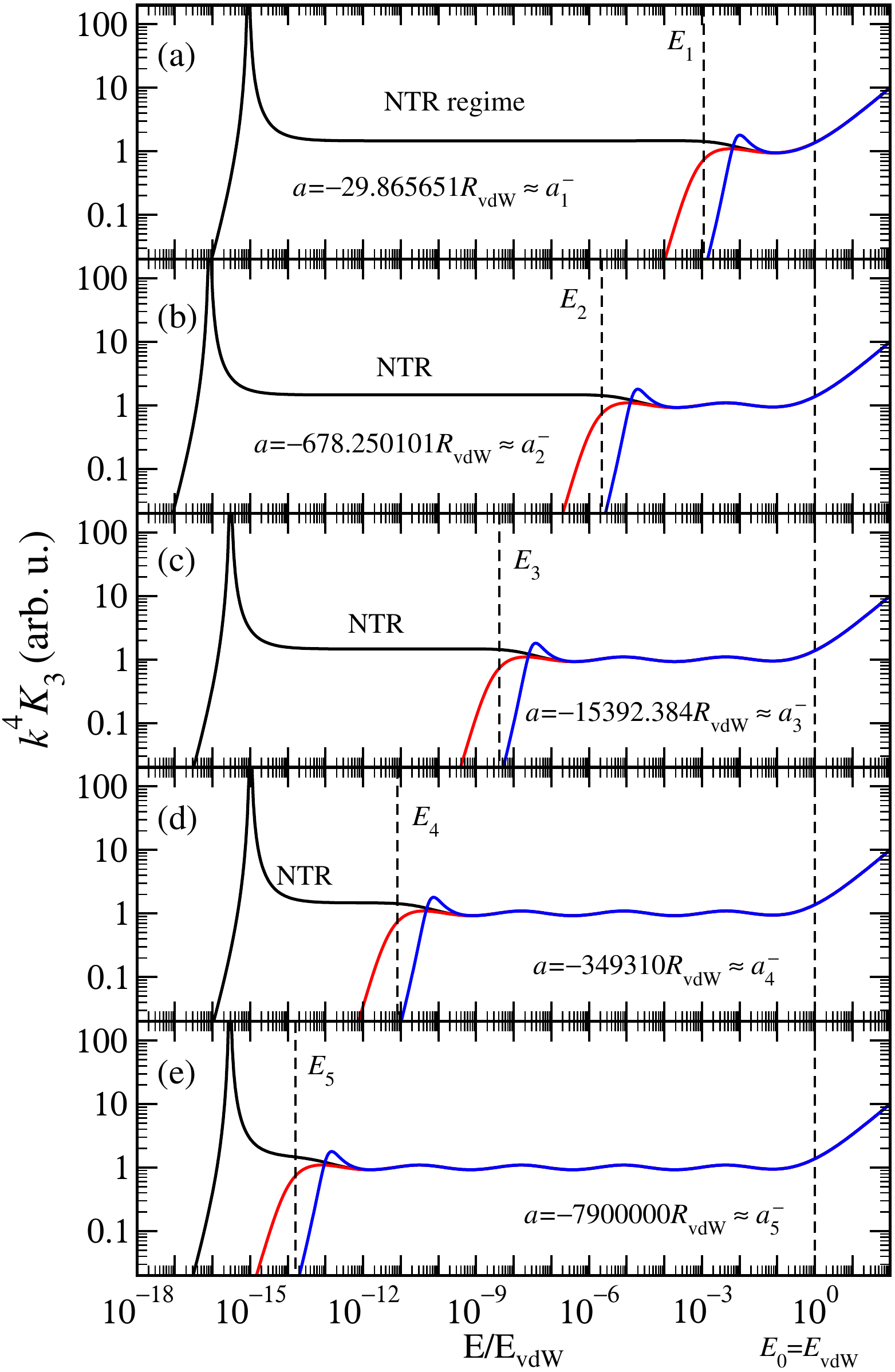}
\caption{\label{single_NTR_5p}  Three-body recombination
  rates for $a<0$. Black curves: resonant cases for the first five
  Efimov states (the values of $a\approx a_n^{-}$ are indicated).  Red
  and blue curves: non-resonant cases corresponding to $a=2a_n^{-}$
  and $a=a_n^{-}/2$ respectively. The dashed vertical line at
  $E_0=E_{\mathrm{vdW}}$ marks the energy scale associated with short
  range physics. The dashed vertical lines at $E_n\equiv
  E_{a_n^{-}}\propto 1/(a_n^{-})^{2}$ mark the boundary between the
  NTR and oscillatory regimes.  }
\end{figure}

\begin{figure}
\includegraphics[width=1.0\columnwidth,clip]{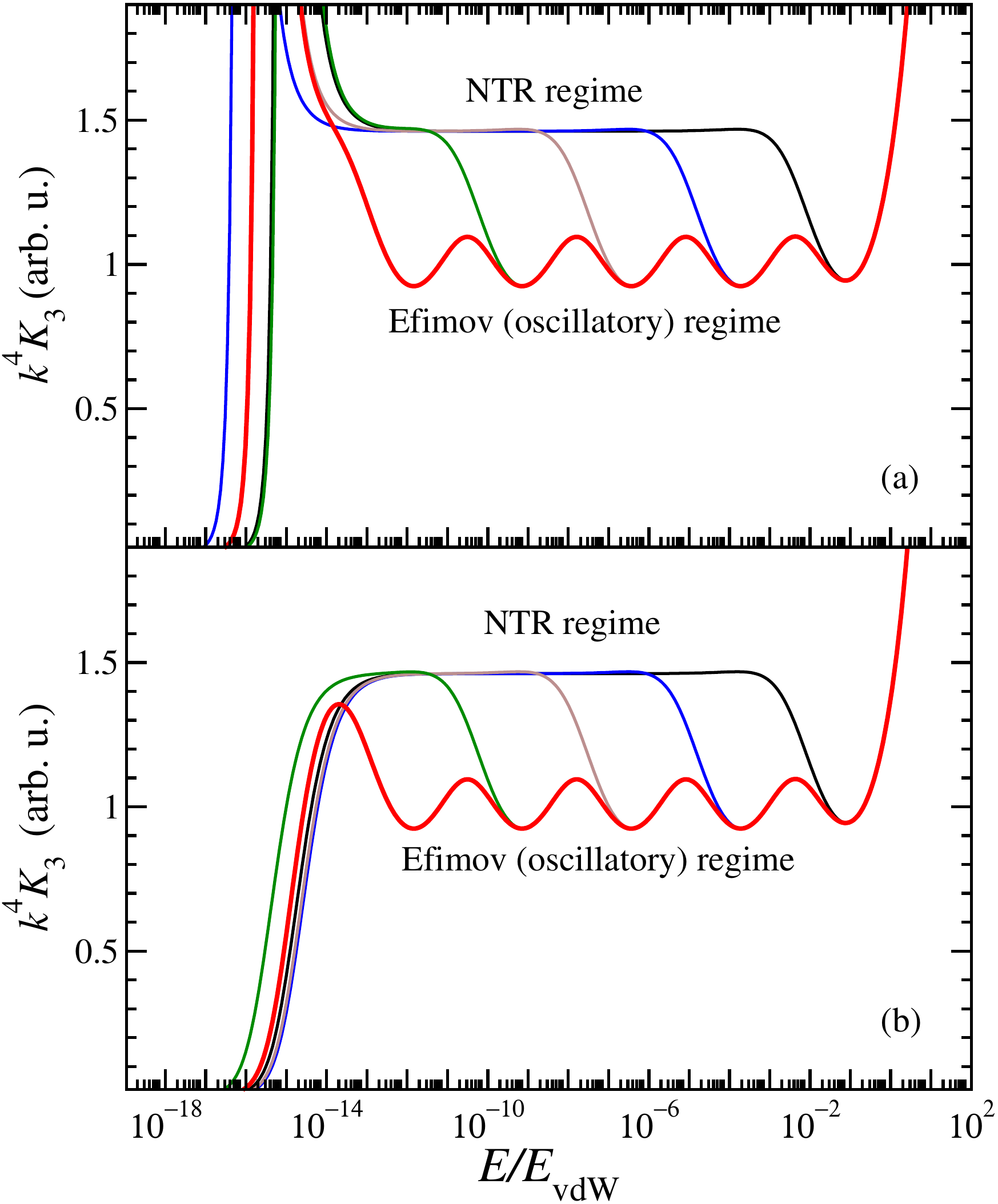}
\caption{\label{single_NTR_2p}  (a) $K_3$ for the case when the Efimov states are shape resonances just above the threshold; these are the black curves from Fig.~\ref{single_NTR_5p}. (b)  $K_3$ for the case when the Efimov states are bound just below the threshold corresponding to values of $|a|$ slightly larger than the values in Fig.~\ref{single_NTR_5p}.}
\end{figure}

Using the model described in Sec.~\ref{theory_1_channel}, we carefully
tune the two-body scattering length $a$ such that there is an Efimov
state extremely close to the threshold.  Here, since there is only one
channel corresponding to the three free atoms, the threshold refers to
it (see Fig.~\ref{nega_a_potential}). In Fig.~\ref{single_NTR_5p}, we
show results for the first five Efimov states.
  Although arbitrary units are used for the
  three-body recombination rate $K_3$ in Fig.~\ref{single_NTR_5p}, one
  could introduce a multiplicative parameter on the right hand side of
 Eqs.~(\ref{Smat_F}) and (\ref{K3_F}) to adjust the
  overall magnitude of $K_3$ to match
  experimental values.   In Fig.~\ref{single_NTR_5p}(a),
  the black curve corresponds to $a_1^-/R_{\mathrm{vdW}}=-29.865651$ and
  displays the first Efimov state as a shape resonance located at very
  low energy, while the red and blue curves correspond to $a=2a_1^-$ and
  $a=a_1^-/2$ respectively. We found that the resonant behavior
  manifests itself only for values of $a$ within one percent of $a_1^-$.  For
  values of $a$ outside of this narrow window, the behavior is similar
  to that shown by the red and blue curves, which we call
  non-resonant.  Although the sharp peak at very low energy is
  striking, this near threshold resonance (NTR) produces a resonant
  enhancement for a much wider energy range. More specifically,
  Fig.~\ref{single_NTR_5p}(a) shows that in the resonant case most of
  the low energy regime is characterized by a new type of behavior;
  namely, $k^4K_3$ is constant for energies between the peak and
  vertical dashed line at $E=E_1$, which we refer to as the NTR
  regime.
For the subsequent Efimov states corresponding to $a_n^-$ with
$n=2,3,4,5$, an oscillatory regime develops \cite{NJP_Wang} in the
energy range $E_n<E<E_0$ between the two vertical dashed lines, as
shown in Fig.~\ref{single_NTR_5p}(b)--(e).   $E_0=E_{\rm vdW}$ denotes
the energy scale associated with short-range physics, while $E_n\equiv
E_{a_n^-}$ with $E_a=(R_{\rm vdW}/a)^2E_{\rm vdW}$  the energy scale
given by the centrifugal barrier near $R=|a|$ in Fig.~1.
 With increasing $|a|$, a
new Efimov state will appear for each critical value $a_n^-$, and we
confirm the well known result $a_{n+1}^-/a_{n}^-=e^{\pi/s_0}$.  Our
  results also confirm that as $|a|$ increases, more oscillations
  appear at lower energies (reflecting the number of bound states),
  while the oscillations at higher energy remain unchanged
  \cite{NJP_Wang}, as shown in Fig.~\ref{single_NTR_5p}, and
  summarized in Fig.~\ref{single_NTR_2p}.

Although the results in Figs.~\ref{single_NTR_5p} and
\ref{single_NTR_2p} were obtained using the smooth potential shown in
Fig.~\ref{nega_a_potential}, the oscillatory behavior can be explained
by writing the wavefunction corresponding to the step potential in
Fig.~\ref{nega_a_potential} in terms of Bessel functions. For $R_0 < R
< |a|$ the regular solution defined in Sec.\ref{theory_1_channel} can
be written as linear combination of Bessel functions of imaginary
order,
\begin{equation}
\begin{array}{l}
\tilde{j}(kR) \equiv \sqrt{\frac{\pi k R}{2}} \mathrm{sech}\left(\frac{\pi s_0}{2}\right) \mathrm{Re}[J_{is_0}(kR)] \\
\tilde{n}(kR) \equiv \sqrt{\frac{\pi k R}{2}} \mathrm{sech}\left(\frac{\pi s_0}{2}\right) \mathrm{Re}[Y_{is_0}(kR)].
\end{array}
\end{equation}
Using the small argument behavior \cite{nist_bessel_hand_book} near  $R \approx R_0$,
\begin{equation}
\begin{array}{l}
\tilde{j}(kR) \sim \left(\frac{t_0kR}{s_0}\right)^{\frac{1}{2}}\cos [s_0 \ln(\textstyle\frac{kR}{2})-\gamma_{s_0}]\\
\tilde{n}(kR) \sim \left(\frac{kR}{t_0s_0}\right)^{\frac{1}{2}}\sin [s_0 \ln(\textstyle\frac{kR}{2})-\gamma_{s_0}],
\end{array}
\end{equation}
where $\gamma_{s_0}=\arg[\Gamma(1+is_0)]$ and $t_0=\tanh (\frac{\pi s_0}{2})$.
Thus the suitable linear combination for the regular solution is
\begin{equation} \label{regular_ugly}
\begin{array}{ll}
\phi_k(R) = &\left(\frac{R_0t_0}{s_0	k}\right)^{\frac{1}{2}} 
     \bigg(\cos [s_0 \ln(\textstyle\frac{kR_0}{2})-\gamma_{s_0}] \tilde{j}(kR) \\
&\makebox[.35in]{ } - t_0^{-1}\sin [s_0 \ln(\textstyle\frac{kR_0}{2})-\gamma_{s_0}]\tilde{n}(kR)\bigg).
\end{array}
\end{equation}
For $R > |a|$, as mentioned in Sec.~\ref{theory_1_channel}, the regular solution can be written as 
a linear combination of the Riccati--Bessel functions
\begin{equation}\label{regular_AB}
k^{\ell_{\mathrm{eff}}+1}\phi_k(R) = A(k)j_{\ell_{\mathrm{eff}}}(kR) + B(k)n_{\ell_{\mathrm{eff}}}(kR).
\end{equation}
By matching at $R=|a|$, i.e., equating the expressions in Eq.~(\ref{regular_ugly}) and Eq.~(\ref{regular_AB}) 
and their derivatives, and using the behavior at large argument \cite{nist_bessel_hand_book} near $R=|a|$,
\begin{equation}
\begin{array}{l}
\tilde{j}(kR) \sim \cos(kR - \textstyle\frac{\pi}{4})\\
\tilde{n}(kR) \sim \sin(kR - \textstyle\frac{\pi}{4})
\end{array}
\end{equation}
we obtain
\begin{equation}
\begin{array}{l}
A(k) =k^2\left(\frac{R_0}{s_0t_0}\right)^{\frac{1}{2}}  \sin [s_0 \ln(\textstyle\frac{kR_0}{2})-\gamma_{s_0}] \\
B(k) = k^2\left(\frac{R_0t_0}{s_0}\right)^{\frac{1}{2}} \cos [s_0 \ln(\textstyle\frac{kR_0}{2})-\gamma_{s_0}], 
\end{array}
\end{equation} 
where $\gamma_{s_0}=\arg[\Gamma(1+is_0)]$ and $t_0=\tanh (\frac{\pi s_0}{2})$.
Therefore, inside the oscillatory regime, the three-body recombination rate reads,
\begin{equation} \label{K3_anege_ugly}
K_3(k) \sim \frac{1}{k^4}\frac{s_0t_0}{R_0}\frac{1}{t_0 + (1-t_0)\sin^2 [s_0 \ln(\textstyle\frac{kR_0}{2})-\gamma_{s_0}]}.
\end{equation}
Note that the amplitude of the oscillatory term, $1-t_0 \approx 0.081$, is much smaller than the background 
term $t_0 \approx 0.92$, which makes it difficult to discern the difference between the NTR regime and the 
oscillatory regime in Fig.~\ref{single_NTR_5p}. 
For $ k \ll |a|^{-1}$, the Jost function can be expanded as a power series, 
$A(k) \sim A_0 + A_2k^2+\mathcal O(k^4)$ and 
$B(k) \sim k^{2\ell_{\rm eff}+1}(B_0 + B_2 k^2 + \dots) =B_0 k^4 +\mathcal O(k^6)$ 
(since $\ell_{\rm eff}=3/2$) \cite{Jost_paper}. Using Eq.~(\ref{K3_F}), we obtained a simple 
expression of the three-body recombination rate, 
\begin{equation}\label{K3_single_anega_lowk}
K_3(k) \sim \frac{1}{(A_0 + A_2k^2)^2 + B_0^2k^8}.
\end{equation}
In the Wigner regime, $A_0$ is dominant, and the zero energy limit of $K_3$ is a constant. 
For the resonant case $A_0$ is very small, which restricts the Wigner regime to the extreme 
ultracold. Thus when $k$ increases, the $A_2k^2$ term quickly becomes dominant and $K_3(k)$ 
reads,
\begin{equation}
K_3(E)  \sim \frac{1}{k^4A_2^2},
\end{equation}
which corresponds to the flat NTR regime shown in
Fig.~\ref{single_NTR_2p} for $k^4K_3$.  Note that the transition
between Wigner and NTR regimes takes place near
$k_{\mathrm{NTR}}=\sqrt{|A_0/A_2|}$, which can be
  estimated from the simple parametrizations $A_0(a)\propto a-a_n^-$
  and $A_2(a) \approx A_2(a_n^-)={\rm const.}$ in the narrow window of
  values of $a$ near $a_n^-$. Fig.~\ref{single_NTR_2p}(a) and
Fig.~\ref{single_NTR_5p} show the Efimov states as shape resonances,
which correspond to the case when $A_0$ and $A_2$ have opposite sign,
such that $A(k)=0$ at $k_{\mathrm{NTR}}$. Note that the very small but
finite term $B_0^2k^8$ in Eq.~(\ref{K3_single_anega_lowk}) will
prevent $K_3$ from diverging.  In Fig.~\ref{single_NTR_2p}(b), the
Efimov states are bound just below the three-body threshold, which
corresponds to the case when $A_0$ and $A_2$ have the same sign.  Thus
$B_0^2k^8$ can be neglected in Eq.~(\ref{K3_single_anega_lowk}), and
the transition between the Wigner and NTR regimes is smooth. We
emphasize that the $K_3 \sim k^{-4}$ behavior in the NTR regime is
accidentally identical to the background $k$-dependence in the
oscillatory regime in Eq.~(\ref{K3_anege_ugly}), though
  the exact values are offset as depicted in
  Fig.~\ref{single_NTR_2p}.
  
\subsection{Single channel results for $a>0$}

When $a>0$, the Efimov potential correlates with the weakly bound
dimer channel, defining the scattering threshold $k=0$.  We expect
that a near threshold Efimov state will strongly affect the Feshbach-molecule--atom collisions, resulting in vibrational relaxation into
deeply bound dimer states.  Here we analyze the vibrational relaxation
rate using the single-channel model described in
Section~\ref{theory_1_channel_relax} , with the vibrational relaxation
rate given by Eq.~(\ref{eq:K_rel}),
\begin{equation}
K_{\mathrm{rel}}(k) \sim \frac{1}{|F(k)|^2}.
\end{equation}
Note that $F(k)$ is the single channel Jost function corresponding to
the lower potential curve in Fig.~\ref{posi_a_potential}.

\begin{figure}
\includegraphics[width=1.0\columnwidth,clip]{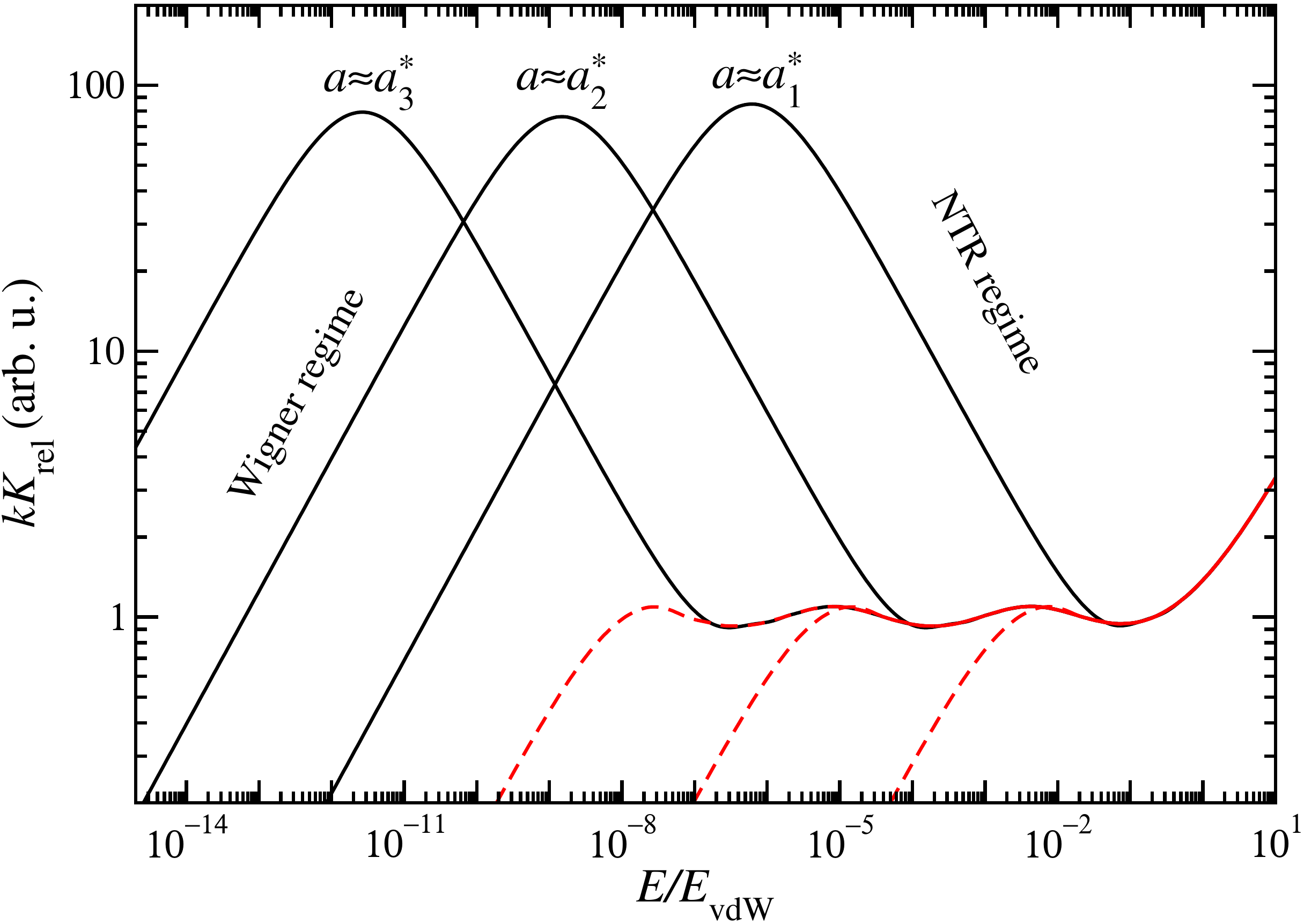}
\caption{\label{NTR_rel} Black curves: vibrational relaxation rate for
  three values of $a=a_n^*$ as indicated, corresponding
  to the first, second and third Efimov state near the threshold,
   with $a_1^*=11.45 \;R_{\rm vdW}$, $a_2^*=260.0\;
    R_{\rm vdW}$, and $a_3^*=5902.0 \;R_{\rm vdW}$, respectively .
  Red curves: relaxation rate for the non-resonant cases
    with $a=3a_n^*$ for each curve, respectively.}
\end{figure}

For energies higher than the binding energy of the shallow dimer
state, $E_b \sim -1/a^2$, but lower than the short range energy scale
$E_0 \sim 1/R_0^2$, i.e., $a^{-1}\ll k \ll R_0^{-1}$,
using Eq.~(\ref{regular_ugly}) again, but this time matching it with
the asymptotic form of $\phi_k(R)$ for $\ell=0$, namely
\begin{equation}
k\phi_k(R) = A(k)\sin (kR) + B(k) \cos (kR) \;,
\end{equation}
we obtain
\begin{equation}
\begin{array}{ll}
A(k) = &\sqrt{k}\left(\frac{R_0t_0}{2s_0}\right)^{\frac{1}{2}}\bigg(\cos [s_0 \ln(\textstyle\frac{kR_0}{2})-\gamma_{s_0}]  \\
&-\textstyle\frac{1}{t_0}\sin [s_0 \ln(\frac{kR_0}{2})-\gamma_{s_0}]\bigg) \\
B(k) =&-\sqrt{k}\left(\frac{R_0t_0}{2s_0}\right)^{\frac{1}{2}}\bigg(\cos [s_0 \ln(\textstyle\frac{kR_0}{2})-\gamma_{s_0}]  \\
&+\textstyle\frac{1}{t_0}\sin [s_0 \ln(\frac{kR_0}{2})-\gamma_{s_0}]\bigg),
\end{array}
\end{equation} 
and thus, $K_{\mathrm{rel}}(k)$ reads,
\begin{equation} \label{Vrel_aposi_ugly}
K_{\mathrm{rel}} \sim \frac{1}{k}\frac{s_0t_0}{R_0}\frac{1}{t_0^2+(1-t_0^2)\sin^2 [s_0 \ln(\frac{kR_0}{2})-\gamma_{s_0}]}.
\end{equation}
The overall $k^{-1}$ behavior of the relaxation rate in
Eq.~(\ref{Vrel_aposi_ugly}) was already mentioned in
Ref.~\cite{NJP_Wang}, where however no oscillatory behavior was found.
In our case, we do obtain an oscillatory behavior for $K_{\rm rel}(k)$
similar to that of $K_3(k)$. As shown in Fig.~\ref{NTR_rel}, this
oscillatory behavior resembles the oscillations in
Fig.~\ref{single_NTR_5p}. The dashed curves in Fig.~\ref{NTR_rel}
correspond to non-resonant cases, while the solid curves show the
resonant cases for the first three Efimov states near the threshold.

For $k \ll a^{-1}$, $A(k)$ and $B(k)$ can be written as a power series: 
$A(k) \sim A_0 + A_2k^2 + \mathcal O (k^4)$ and 
$B(k) = k^{2\ell +1}(B_0 +B_2k^2 +\dots) \sim B_0 k + \mathcal O (k^3)$ (with $\ell =0)$. 
As a result the relaxation rate reads
\begin{equation}\label{Vrel_single_aposi_lowk}
K_{\mathrm{rel}}(k) \sim \frac{1}{(A_0+A_2k^2)^2  + B_0^2k^2}.
\end{equation}
$A_0$ is very small for the resonant case and thus the competition
between $A_0$ and $B_0^2k^2$ gives the Wigner and NTR regimes. In the
Wigner regime, $A_0$ is dominant and $K_{\mathrm{rel}}$ is a constant
when $k$ goes to zero, while in the NTR regime $K_{\mathrm{rel}} \sim
1/B_0^2k^2$ since $B_0^2k^2$ becomes dominant in the denominator of
Eq.~(\ref{Vrel_single_aposi_lowk}) as we increase $k$. In contrast to
Eq.~(\ref{K3_single_anega_lowk}), the $A_2k^2$ term no longer plays a
significant role and we omit it from
Eq.~(\ref{Vrel_single_aposi_lowk}); indeed, $A_2^2k^4$ is
  a higher order term which can be neglected, while the cross term
  $2A_0A_2k^2$ can be combined with $B_0^2k^2$ which is equivalent to
  altering $B_0$ very slightly.  Hence we obtain
\begin{equation}
   K_{\mathrm{rel}}(k) \sim \frac{1}{A_0^2  + B_0^2k^2}.
\end{equation}
This equation captures the transition between the Wigner and NTR
regimes. The transition is smooth whether or not the Efimov state is
just above or below the threshold.

\subsection{Two channel results for $a>0$}
\label{2_channel}

\begin{figure}
\includegraphics[width=1\columnwidth,clip]{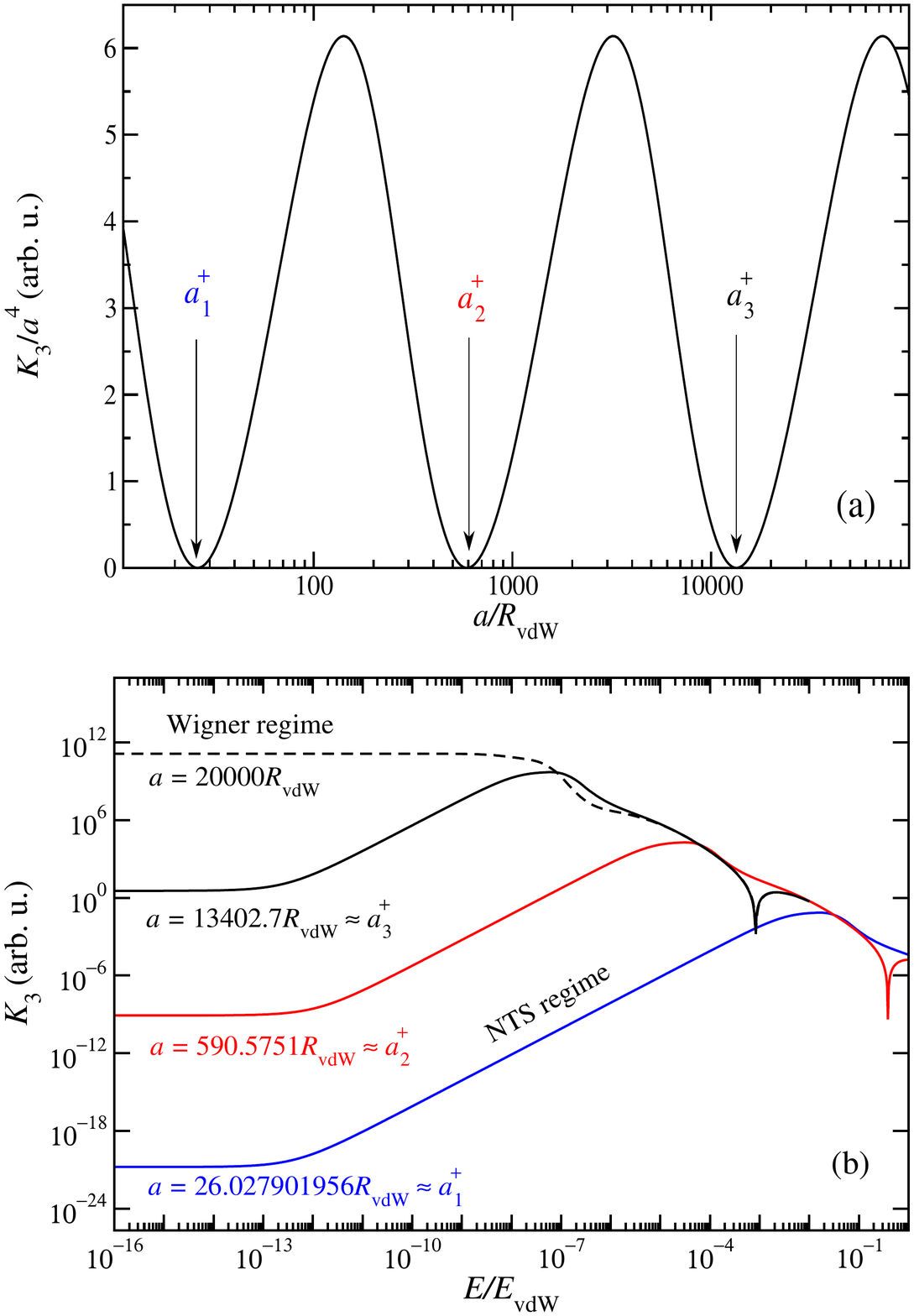}
\caption{\label{NTS} (a) $K_3$ in the limit $E \rightarrow
  0$ as function of $a$. The positions $a_n^+$ of the
  minima are indicated and their values are given in the lower panel.
  (b) Dashed line: $K_3(E)$ for non-critical case, which displays only
  the Wigner regime behavior at low energy. Solid lines: first three
  critical cases corresponding to the minima in the upper panel. The
  minima of $K_3(E)$ appearing at energies between $E/E_{\mathrm{vdW}}
  = 10^{-4}$ and $E/E_{\mathrm{vdW}} =1$ are due to destructive
  interference as explained in Ref.~\cite{PRL104_Wang}.  }
\end{figure}

When the atom-atom scattering length $a$ is positive and large, there
is a shallow dimer state just below the threshold, and we use the
2-channel model introduced in Sec.~\ref{theory_2_channel} to compute
the three-body recombination rate for the process $B+B+B \to B + B_2
(\mathrm{shallow})$. As is well known \cite{Physrep428_Braaten}, the
zero energy limit of the three-body recombination rate $K_3$ is a
log-periodic function of the dimer scattering length $a$, as shown in
Fig.~\ref{NTS}a. Each maximum in Fig.~\ref{NTS}a corresponds to an
Efimov state at the dimer-atom threshold \cite{Physrep428_Braaten},
which does not affect the energy dependence of $K_3$. However, each
minimum in Fig.~\ref{NTS}a corresponds to a critical value of $a$ for
which the energy dependence of $K_3$ is dramatically modified as shown
in Fig.~\ref{NTS}b. The dashed line in Fig.~\ref{NTS}b corresponds to
a non-critical case, when $K_3(k)$ follows Wigner's threshold law for
$k \ll a^{-1}$. The other three curves in Fig.~\ref{NTS}b correspond
to the three minima in Fig.~\ref{NTS}a, for which the three-body
recombination rate displays a strongly suppressed behavior $K_3(k)
\sim k^4$ for $k_{\mathrm{NTS}} \ll k \ll a^{-1}$, that we label
the {\it Near Threshold Suppression} (or NTS) regime.

The numerical solution for the 2-channel model is obtained as follows.
The regular matrix solution $\Phi$ is initialized at $R=R_0$ using
boundary conditions corresponding to a hard wall, $\Phi(R_0)=0$ and
$\Phi'(R_0) = \mathbf{I}$, where $\mathbf{I}$ is the $2 \times 2$ unit
matrix. After propagation, and recalling that $\ell = \ell_{\rm eff}$
for channel 1, and $\ell=0$ for channel 2, the regular solution is
matched to asymptotic solutions,
\begin{equation}
\Phi \mathbf{k}^{\ell+1} = \mathbf{fA + gB},
\end{equation}
where $\mathbf{k}^{\ell+1}=\mathrm{diag}(k_1^{\ell_{\mathrm{eff}}+1},k_2)$ and $\mathbf{f}$ and $\mathbf{g}$ are diagonal matrices containing the single channel asymptotic solutions
$f_{1} = j_{\ell_{\mathrm{eff}}}(k_1R)$, $g_{1} = n_{\ell_{\mathrm{eff}}}(k_1R)$, $f_{2} = \sin (k_2 R)$ and $g_2 = \cos(k_2 R)$. The matrices $\mathbf{A}$ and $\mathbf{B}$ are the real and imaginary parts of the Jost matrix $\mathbf{F}=\mathbf{A} - i\mathbf{B}$. The expression of the S-matrix in terms of the Jost matrix reads
\begin{equation}
S_{21} = \sqrt{\frac{k_2}{k_1}}(\mathbf{F^{*}F}^{-1})_{21} = 2i\sqrt{\frac{k_2}{k_1}}\frac{A_{22}B_{21}-A_{21}B_{22}}{\det (\mathbf{F})}.
\end{equation}
Using the power series of the Jost matrix elements when $k \ll a^{-1}$, we obtain
\begin{equation}\label{S21}
S_{21}(k_1) = k_1^2\frac{C_0+C_2k_1^2}{D_0+D_2k_1^2}.
\end{equation}
Recalling Eq.~(\ref{K3_F}), we find
\begin{equation}
   K_3 \sim  \frac{1}{k_1^4}|S_{21}|^2 = \left| \frac{C_0+C_2k_1^2}{D_0+D_2k_1^2} \right|^2 \;.
   \label{eq:K3_21}
\end{equation}
For the values of $a$ corresponding to the strongly suppressed NTS cases shown in Fig.~\ref{NTS}b,
the Efimov states are not near the threshold. This implies that the S-matrix does not exhibit a resonant
structure, i.e., it has no nearby pole, or equivalently, $\det (\mathbf{F})\neq 0$ or simply $D_0\neq 0$
in the limit $k_1\rightarrow 0$. Hence the low $k_1$ behavior of the three-body recombination rate
$K_3$ is determined by the competition between $C_0$ and $C_2k_1^2$
in Eq.~(\ref{eq:K3_21}). In the Wigner regime, $C_0$ is dominant and thus $K_3$ is nearly constant.
Note that normally, the Wigner regime behavior is valid for $k_1 \ll a^{-1}$, while in the NTS case $C_0$
is vanishingly small and the Wigner regime is restricted to
$k_1 \ll k_{\mathrm{NTS}} = \sqrt{|C_0/C_2|}$. As Fig.~\ref{NTS}b clearly shows, there is a new
type of behavior for $k_{\mathrm{NTS}} \ll k_1 \ll a^{-1}$, where $C_2k_1^2$ in Eq.~(\ref{S21}) is
dominant, and thus $K_3 \sim k_1^4$.

\section{Conclusion}

In this paper, we studied the effect of Efimov states on the energy
dependence of three-body recombination rates $K_3(E)$ and vibrational
relaxation rates $K_{\mathrm{rel}}(E)$ for shallow dimers. We used
simple models capturing the essential physical processes. For negative
atom-atom scattering length, $a<0$, a single-channel model is used to
expressed $K_3$ in term of a single-channel Jost function. For
positive scattering length, $a>0$, we use a single-channel model
expressing the vibrational relaxation $K_{\rm rel}$ in term of a
single-channel Jost function, while a two-channel model is used to
obtain $K_3$. We used the analytical structure and properties of Jost
functions to explain the numerical results obtained by varying the
scattering energy $E$ (or momentum $k$) for given values of $a$.

When the atom-atom scattering length $a$ is negative, we
recovered the usual Wigner regime at ultralow energies, but as $a$
gets very close to an Efimov resonance, we uncovered a new regime,
labeled Near Threshold Resonance (or NTR) regime where $k^4 K_3\sim
\mbox{const.}$ (or low energy behavior $K_3(E)\sim E^{-2}$). We also
recover the oscillatory behavior obtained in other studies. Similarly,
when $a \to +\infty$, we find the relaxation rate to reach a constant
at ultralow energy, corresponding to the Wigner regime, and if an
Efimov state is at the threshold, $K_{\mathrm{rel}}\sim k^{-2} \propto
E^{-1}$, corresponding again to the NTR regime. In
  addition, we found oscillations about a $k^{-1}$ behavior for $k$
  corresponding to energies larger than the NTR regime.

Finally, for $a \to +\infty$, we found that the usual Wigner behavior
$K_3\sim \mbox{const.}$ for the rate of three-body recombination (into
a shallow dimer) is strongly suppressed near the
threshold.  It behaves as $k^{4}$ or $K_3(E)\sim E^{2}$
when the atom-atom scattering length nears one of the critical values
corresponding to the minima of $K_3$ as a function of $a$
(in the zero energy limit).  This new regime, labeled the
Near Threshold Suppression regime (NTS), could potentially be used to
stabilize cold gases by preventing three-body recombination by
adjusting the two-body scattering length to one of  the
  large and positive critical values of $a$.

\section{Acknowledgment}

The authors wish to thank Dr.~Jia Wang for fruitful discussions. This work was partially supported
by the US Army Research Office, Chemistry Division, Grant No.~W911NF-13-1-0213 (DS), and by
the MURI US Army Research Office Grant No.~W911NF-14-1-0378 (IS and RC).

\bibliography{Efimov}

\begin{thebibliography}{42}
\expandafter\ifx\csname natexlab\endcsname\relax\def\natexlab#1{#1}\fi
\expandafter\ifx\csname bibnamefont\endcsname\relax
  \def\bibnamefont#1{#1}\fi
\expandafter\ifx\csname bibfnamefont\endcsname\relax
  \def\bibfnamefont#1{#1}\fi
\expandafter\ifx\csname citenamefont\endcsname\relax
  \def\citenamefont#1{#1}\fi
\expandafter\ifx\csname url\endcsname\relax
  \def\url#1{\texttt{#1}}\fi
\expandafter\ifx\csname urlprefix\endcsname\relax\def\urlprefix{URL }\fi
\providecommand{\bibinfo}[2]{#2}
\providecommand{\eprint}[2][]{\url{#2}}

\bibitem[{\citenamefont{Neirotti et~al.}(1997)\citenamefont{Neirotti, Serra,
  and Kais}}]{Kais1}
\bibinfo{author}{\bibfnamefont{J.~P.} \bibnamefont{Neirotti}},
  \bibinfo{author}{\bibfnamefont{P.}~\bibnamefont{Serra}}, \bibnamefont{and}
  \bibinfo{author}{\bibfnamefont{S.}~\bibnamefont{Kais}},
  \bibinfo{journal}{Phys. Rev. Lett.} \textbf{\bibinfo{volume}{79}},
  \bibinfo{pages}{3142} (\bibinfo{year}{1997}),
  \urlprefix\url{http://link.aps.org/doi/10.1103/PhysRevLett.79.3142}.

\bibitem[{\citenamefont{Serra et~al.}(1998)\citenamefont{Serra, Neirotti, and
  Kais}}]{Kais2}
\bibinfo{author}{\bibfnamefont{P.}~\bibnamefont{Serra}},
  \bibinfo{author}{\bibfnamefont{J.~P.} \bibnamefont{Neirotti}},
  \bibnamefont{and} \bibinfo{author}{\bibfnamefont{S.}~\bibnamefont{Kais}},
  \bibinfo{journal}{Phys. Rev. Lett.} \textbf{\bibinfo{volume}{80}},
  \bibinfo{pages}{5293} (\bibinfo{year}{1998}),
  \urlprefix\url{http://link.aps.org/doi/10.1103/PhysRevLett.80.5293}.

\bibitem[{\citenamefont{Kais and Shi}(2000)}]{Kais3}
\bibinfo{author}{\bibfnamefont{S.}~\bibnamefont{Kais}} \bibnamefont{and}
  \bibinfo{author}{\bibfnamefont{Q.}~\bibnamefont{Shi}},
  \bibinfo{journal}{Phys. Rev. A} \textbf{\bibinfo{volume}{62}},
  \bibinfo{pages}{060502} (\bibinfo{year}{2000}),
  \urlprefix\url{http://link.aps.org/doi/10.1103/PhysRevA.62.060502}.

\bibitem[{\citenamefont{Efimov}(1991)}]{nuc1}
\bibinfo{author}{\bibfnamefont{V.}~\bibnamefont{Efimov}},
  \bibinfo{journal}{Phys. Rev. C} \textbf{\bibinfo{volume}{44}},
  \bibinfo{pages}{2303} (\bibinfo{year}{1991}),
  \urlprefix\url{http://link.aps.org/doi/10.1103/PhysRevC.44.2303}.

\bibitem[{\citenamefont{Efimov}(1993)}]{nuc2}
\bibinfo{author}{\bibfnamefont{V.}~\bibnamefont{Efimov}},
  \bibinfo{journal}{Phys. Rev. C} \textbf{\bibinfo{volume}{47}},
  \bibinfo{pages}{1876} (\bibinfo{year}{1993}),
  \urlprefix\url{http://link.aps.org/doi/10.1103/PhysRevC.47.1876}.

\bibitem[{\citenamefont{Garrido et~al.}(2006)\citenamefont{Garrido, Fedorov,
  and Jensen}}]{nuc3}
\bibinfo{author}{\bibfnamefont{E.}~\bibnamefont{Garrido}},
  \bibinfo{author}{\bibfnamefont{D.~V.} \bibnamefont{Fedorov}},
  \bibnamefont{and} \bibinfo{author}{\bibfnamefont{A.~S.}
  \bibnamefont{Jensen}}, \bibinfo{journal}{Phys. Rev. Lett.}
  \textbf{\bibinfo{volume}{96}}, \bibinfo{pages}{112501}
  (\bibinfo{year}{2006}),
  \urlprefix\url{http://link.aps.org/doi/10.1103/PhysRevLett.96.112501}.

\bibitem[{\citenamefont{Efimov}(1971)}]{Efimov_original}
\bibinfo{author}{\bibfnamefont{V.}~\bibnamefont{Efimov}},
  \bibinfo{journal}{Soviet Journal of Nuclear Physics}
  \textbf{\bibinfo{volume}{12}} (\bibinfo{year}{1971}).

\bibitem[{\citenamefont{Efimov}(1970)}]{Efimov_original2}
\bibinfo{author}{\bibfnamefont{V.}~\bibnamefont{Efimov}},
  \bibinfo{journal}{Physics Letters B} \textbf{\bibinfo{volume}{33}},
  \bibinfo{pages}{563 } (\bibinfo{year}{1970}), ISSN \bibinfo{issn}{0370-2693},
  \urlprefix\url{http://www.sciencedirect.com/science/article/pii/0370269370903497}.

\bibitem[{\citenamefont{Braaten and Hammer}(2006)}]{Physrep428_Braaten}
\bibinfo{author}{\bibfnamefont{E.}~\bibnamefont{Braaten}} \bibnamefont{and}
  \bibinfo{author}{\bibfnamefont{H.-W.} \bibnamefont{Hammer}},
  \bibinfo{journal}{Physics Reports} \textbf{\bibinfo{volume}{428}},
  \bibinfo{pages}{259 } (\bibinfo{year}{2006}), ISSN \bibinfo{issn}{0370-1573},
  \urlprefix\url{http://www.sciencedirect.com/science/article/pii/S0370157306000822}.

\bibitem[{\citenamefont{Kraemer et~al.}(2006)\citenamefont{Kraemer, Mark,
  Waldburger, Danzl, Chin, Engeser, Lange, Pilch, Jaakkola, N{\"a}gerl
  et~al.}}]{Nature440_315}
\bibinfo{author}{\bibfnamefont{T.}~\bibnamefont{Kraemer}},
  \bibinfo{author}{\bibfnamefont{M.}~\bibnamefont{Mark}},
  \bibinfo{author}{\bibfnamefont{P.}~\bibnamefont{Waldburger}},
  \bibinfo{author}{\bibfnamefont{J.~G.} \bibnamefont{Danzl}},
  \bibinfo{author}{\bibfnamefont{C.}~\bibnamefont{Chin}},
  \bibinfo{author}{\bibfnamefont{B.}~\bibnamefont{Engeser}},
  \bibinfo{author}{\bibfnamefont{A.~D.} \bibnamefont{Lange}},
  \bibinfo{author}{\bibfnamefont{K.}~\bibnamefont{Pilch}},
  \bibinfo{author}{\bibfnamefont{A.}~\bibnamefont{Jaakkola}},
  \bibinfo{author}{\bibfnamefont{H.-C.} \bibnamefont{N{\"a}gerl}},
  \bibnamefont{et~al.}, \bibinfo{journal}{Nature}
  \textbf{\bibinfo{volume}{440}}, \bibinfo{pages}{315} (\bibinfo{year}{2006}).

\bibitem[{\citenamefont{Williams et~al.}(2009)\citenamefont{Williams, Hazlett,
  Huckans, Stites, Zhang, and O'Hara}}]{PRL103_JRwill}
\bibinfo{author}{\bibfnamefont{J.~R.} \bibnamefont{Williams}},
  \bibinfo{author}{\bibfnamefont{E.~L.} \bibnamefont{Hazlett}},
  \bibinfo{author}{\bibfnamefont{J.~H.} \bibnamefont{Huckans}},
  \bibinfo{author}{\bibfnamefont{R.~W.} \bibnamefont{Stites}},
  \bibinfo{author}{\bibfnamefont{Y.}~\bibnamefont{Zhang}}, \bibnamefont{and}
  \bibinfo{author}{\bibfnamefont{K.~M.} \bibnamefont{O'Hara}},
  \bibinfo{journal}{Phys. Rev. Lett.} \textbf{\bibinfo{volume}{103}},
  \bibinfo{pages}{130404} (\bibinfo{year}{2009}),
  \urlprefix\url{http://link.aps.org/doi/10.1103/PhysRevLett.103.130404}.

\bibitem[{\citenamefont{Pollack et~al.}(2009)\citenamefont{Pollack, Dries, and
  Hulet}}]{Science326_Pollack}
\bibinfo{author}{\bibfnamefont{S.~E.} \bibnamefont{Pollack}},
  \bibinfo{author}{\bibfnamefont{D.}~\bibnamefont{Dries}}, \bibnamefont{and}
  \bibinfo{author}{\bibfnamefont{R.~G.} \bibnamefont{Hulet}},
  \bibinfo{journal}{Science} \textbf{\bibinfo{volume}{326}},
  \bibinfo{pages}{1683} (\bibinfo{year}{2009}).

\bibitem[{\citenamefont{Gross et~al.}(2010)\citenamefont{Gross, Shotan,
  Kokkelmans, and Khaykovich}}]{PRL105_Gross}
\bibinfo{author}{\bibfnamefont{N.}~\bibnamefont{Gross}},
  \bibinfo{author}{\bibfnamefont{Z.}~\bibnamefont{Shotan}},
  \bibinfo{author}{\bibfnamefont{S.}~\bibnamefont{Kokkelmans}},
  \bibnamefont{and}
  \bibinfo{author}{\bibfnamefont{L.}~\bibnamefont{Khaykovich}},
  \bibinfo{journal}{Phys. Rev. Lett.} \textbf{\bibinfo{volume}{105}},
  \bibinfo{pages}{103203} (\bibinfo{year}{2010}),
  \urlprefix\url{http://link.aps.org/doi/10.1103/PhysRevLett.105.103203}.

\bibitem[{\citenamefont{Berninger et~al.}(2011)\citenamefont{Berninger,
  Zenesini, Huang, Harm, N\"agerl, Ferlaino, Grimm, Julienne, and
  Hutson}}]{PRL107_Berninger}
\bibinfo{author}{\bibfnamefont{M.}~\bibnamefont{Berninger}},
  \bibinfo{author}{\bibfnamefont{A.}~\bibnamefont{Zenesini}},
  \bibinfo{author}{\bibfnamefont{B.}~\bibnamefont{Huang}},
  \bibinfo{author}{\bibfnamefont{W.}~\bibnamefont{Harm}},
  \bibinfo{author}{\bibfnamefont{H.-C.} \bibnamefont{N\"agerl}},
  \bibinfo{author}{\bibfnamefont{F.}~\bibnamefont{Ferlaino}},
  \bibinfo{author}{\bibfnamefont{R.}~\bibnamefont{Grimm}},
  \bibinfo{author}{\bibfnamefont{P.~S.} \bibnamefont{Julienne}},
  \bibnamefont{and} \bibinfo{author}{\bibfnamefont{J.~M.}
  \bibnamefont{Hutson}}, \bibinfo{journal}{Phys. Rev. Lett.}
  \textbf{\bibinfo{volume}{107}}, \bibinfo{pages}{120401}
  (\bibinfo{year}{2011}),
  \urlprefix\url{http://link.aps.org/doi/10.1103/PhysRevLett.107.120401}.

\bibitem[{\citenamefont{Wild et~al.}(2012)\citenamefont{Wild, Makotyn, Pino,
  Cornell, and Jin}}]{PRL108_Wild}
\bibinfo{author}{\bibfnamefont{R.~J.} \bibnamefont{Wild}},
  \bibinfo{author}{\bibfnamefont{P.}~\bibnamefont{Makotyn}},
  \bibinfo{author}{\bibfnamefont{J.~M.} \bibnamefont{Pino}},
  \bibinfo{author}{\bibfnamefont{E.~A.} \bibnamefont{Cornell}},
  \bibnamefont{and} \bibinfo{author}{\bibfnamefont{D.~S.} \bibnamefont{Jin}},
  \bibinfo{journal}{Phys. Rev. Lett.} \textbf{\bibinfo{volume}{108}},
  \bibinfo{pages}{145305} (\bibinfo{year}{2012}),
  \urlprefix\url{http://link.aps.org/doi/10.1103/PhysRevLett.108.145305}.

\bibitem[{\citenamefont{Roy et~al.}(2013)\citenamefont{Roy, Landini,
  Trenkwalder, Semeghini, Spagnolli, Simoni, Fattori, Inguscio, and
  Modugno}}]{PRL111_Roy}
\bibinfo{author}{\bibfnamefont{S.}~\bibnamefont{Roy}},
  \bibinfo{author}{\bibfnamefont{M.}~\bibnamefont{Landini}},
  \bibinfo{author}{\bibfnamefont{A.}~\bibnamefont{Trenkwalder}},
  \bibinfo{author}{\bibfnamefont{G.}~\bibnamefont{Semeghini}},
  \bibinfo{author}{\bibfnamefont{G.}~\bibnamefont{Spagnolli}},
  \bibinfo{author}{\bibfnamefont{A.}~\bibnamefont{Simoni}},
  \bibinfo{author}{\bibfnamefont{M.}~\bibnamefont{Fattori}},
  \bibinfo{author}{\bibfnamefont{M.}~\bibnamefont{Inguscio}}, \bibnamefont{and}
  \bibinfo{author}{\bibfnamefont{G.}~\bibnamefont{Modugno}},
  \bibinfo{journal}{Phys. Rev. Lett.} \textbf{\bibinfo{volume}{111}},
  \bibinfo{pages}{053202} (\bibinfo{year}{2013}),
  \urlprefix\url{http://link.aps.org/doi/10.1103/PhysRevLett.111.053202}.

\bibitem[{\citenamefont{Ottenstein et~al.}(2008)\citenamefont{Ottenstein,
  Lompe, Kohnen, Wenz, and Jochim}}]{PRL101_Ottenstein}
\bibinfo{author}{\bibfnamefont{T.~B.} \bibnamefont{Ottenstein}},
  \bibinfo{author}{\bibfnamefont{T.}~\bibnamefont{Lompe}},
  \bibinfo{author}{\bibfnamefont{M.}~\bibnamefont{Kohnen}},
  \bibinfo{author}{\bibfnamefont{A.~N.} \bibnamefont{Wenz}}, \bibnamefont{and}
  \bibinfo{author}{\bibfnamefont{S.}~\bibnamefont{Jochim}},
  \bibinfo{journal}{Phys. Rev. Lett.} \textbf{\bibinfo{volume}{101}},
  \bibinfo{pages}{203202} (\bibinfo{year}{2008}),
  \urlprefix\url{http://link.aps.org/doi/10.1103/PhysRevLett.101.203202}.

\bibitem[{\citenamefont{Huckans et~al.}(2009)\citenamefont{Huckans, Williams,
  Hazlett, Stites, and O'Hara}}]{PRL102_Huckans}
\bibinfo{author}{\bibfnamefont{J.~H.} \bibnamefont{Huckans}},
  \bibinfo{author}{\bibfnamefont{J.~R.} \bibnamefont{Williams}},
  \bibinfo{author}{\bibfnamefont{E.~L.} \bibnamefont{Hazlett}},
  \bibinfo{author}{\bibfnamefont{R.~W.} \bibnamefont{Stites}},
  \bibnamefont{and} \bibinfo{author}{\bibfnamefont{K.~M.}
  \bibnamefont{O'Hara}}, \bibinfo{journal}{Phys. Rev. Lett.}
  \textbf{\bibinfo{volume}{102}}, \bibinfo{pages}{165302}
  (\bibinfo{year}{2009}),
  \urlprefix\url{http://link.aps.org/doi/10.1103/PhysRevLett.102.165302}.

\bibitem[{\citenamefont{Nakajima et~al.}(2010)\citenamefont{Nakajima,
  Horikoshi, Mukaiyama, Naidon, and Ueda}}]{PRL105_Nakajima}
\bibinfo{author}{\bibfnamefont{S.}~\bibnamefont{Nakajima}},
  \bibinfo{author}{\bibfnamefont{M.}~\bibnamefont{Horikoshi}},
  \bibinfo{author}{\bibfnamefont{T.}~\bibnamefont{Mukaiyama}},
  \bibinfo{author}{\bibfnamefont{P.}~\bibnamefont{Naidon}}, \bibnamefont{and}
  \bibinfo{author}{\bibfnamefont{M.}~\bibnamefont{Ueda}},
  \bibinfo{journal}{Phys. Rev. Lett.} \textbf{\bibinfo{volume}{105}},
  \bibinfo{pages}{023201} (\bibinfo{year}{2010}),
  \urlprefix\url{http://link.aps.org/doi/10.1103/PhysRevLett.105.023201}.

\bibitem[{\citenamefont{Barontini et~al.}(2009)\citenamefont{Barontini, Weber,
  Rabatti, Catani, Thalhammer, Inguscio, and Minardi}}]{PRL103_Barontini}
\bibinfo{author}{\bibfnamefont{G.}~\bibnamefont{Barontini}},
  \bibinfo{author}{\bibfnamefont{C.}~\bibnamefont{Weber}},
  \bibinfo{author}{\bibfnamefont{F.}~\bibnamefont{Rabatti}},
  \bibinfo{author}{\bibfnamefont{J.}~\bibnamefont{Catani}},
  \bibinfo{author}{\bibfnamefont{G.}~\bibnamefont{Thalhammer}},
  \bibinfo{author}{\bibfnamefont{M.}~\bibnamefont{Inguscio}}, \bibnamefont{and}
  \bibinfo{author}{\bibfnamefont{F.}~\bibnamefont{Minardi}},
  \bibinfo{journal}{Phys. Rev. Lett.} \textbf{\bibinfo{volume}{103}},
  \bibinfo{pages}{043201} (\bibinfo{year}{2009}),
  \urlprefix\url{http://link.aps.org/doi/10.1103/PhysRevLett.103.043201}.

\bibitem[{\citenamefont{Bloom et~al.}(2013)\citenamefont{Bloom, Hu, Cumby, and
  Jin}}]{PRL111_Bloom}
\bibinfo{author}{\bibfnamefont{R.~S.} \bibnamefont{Bloom}},
  \bibinfo{author}{\bibfnamefont{M.-G.} \bibnamefont{Hu}},
  \bibinfo{author}{\bibfnamefont{T.~D.} \bibnamefont{Cumby}}, \bibnamefont{and}
  \bibinfo{author}{\bibfnamefont{D.~S.} \bibnamefont{Jin}},
  \bibinfo{journal}{Phys. Rev. Lett.} \textbf{\bibinfo{volume}{111}},
  \bibinfo{pages}{105301} (\bibinfo{year}{2013}),
  \urlprefix\url{http://link.aps.org/doi/10.1103/PhysRevLett.111.105301}.

\bibitem[{\citenamefont{Pires et~al.}(2014)\citenamefont{Pires, Ulmanis,
  H\"afner, Repp, Arias, Kuhnle, and Weidem\"uller}}]{PRL112_Pires}
\bibinfo{author}{\bibfnamefont{R.}~\bibnamefont{Pires}},
  \bibinfo{author}{\bibfnamefont{J.}~\bibnamefont{Ulmanis}},
  \bibinfo{author}{\bibfnamefont{S.}~\bibnamefont{H\"afner}},
  \bibinfo{author}{\bibfnamefont{M.}~\bibnamefont{Repp}},
  \bibinfo{author}{\bibfnamefont{A.}~\bibnamefont{Arias}},
  \bibinfo{author}{\bibfnamefont{E.~D.} \bibnamefont{Kuhnle}},
  \bibnamefont{and}
  \bibinfo{author}{\bibfnamefont{M.}~\bibnamefont{Weidem\"uller}},
  \bibinfo{journal}{Phys. Rev. Lett.} \textbf{\bibinfo{volume}{112}},
  \bibinfo{pages}{250404} (\bibinfo{year}{2014}),
  \urlprefix\url{http://link.aps.org/doi/10.1103/PhysRevLett.112.250404}.

\bibitem[{\citenamefont{Tung et~al.}(2014)\citenamefont{Tung,
  Jim\'enez-Garc\'{\i}a, Johansen, Parker, and Chin}}]{PRL113_Tung}
\bibinfo{author}{\bibfnamefont{S.-K.} \bibnamefont{Tung}},
  \bibinfo{author}{\bibfnamefont{K.}~\bibnamefont{Jim\'enez-Garc\'{\i}a}},
  \bibinfo{author}{\bibfnamefont{J.}~\bibnamefont{Johansen}},
  \bibinfo{author}{\bibfnamefont{C.~V.} \bibnamefont{Parker}},
  \bibnamefont{and} \bibinfo{author}{\bibfnamefont{C.}~\bibnamefont{Chin}},
  \bibinfo{journal}{Phys. Rev. Lett.} \textbf{\bibinfo{volume}{113}},
  \bibinfo{pages}{240402} (\bibinfo{year}{2014}),
  \urlprefix\url{http://link.aps.org/doi/10.1103/PhysRevLett.113.240402}.

\bibitem[{\citenamefont{Huang et~al.}(2014)\citenamefont{Huang, Sidorenkov,
  Grimm, and Hutson}}]{PRL112_Huang}
\bibinfo{author}{\bibfnamefont{B.}~\bibnamefont{Huang}},
  \bibinfo{author}{\bibfnamefont{L.~A.} \bibnamefont{Sidorenkov}},
  \bibinfo{author}{\bibfnamefont{R.}~\bibnamefont{Grimm}}, \bibnamefont{and}
  \bibinfo{author}{\bibfnamefont{J.~M.} \bibnamefont{Hutson}},
  \bibinfo{journal}{Phys. Rev. Lett.} \textbf{\bibinfo{volume}{112}},
  \bibinfo{pages}{190401} (\bibinfo{year}{2014}),
  \urlprefix\url{http://link.aps.org/doi/10.1103/PhysRevLett.112.190401}.

\bibitem[{\citenamefont{Lompe et~al.}(2010)\citenamefont{Lompe, Ottenstein,
  Serwane, Viering, Wenz, Z\"urn, and Jochim}}]{PRL105_Lompe}
\bibinfo{author}{\bibfnamefont{T.}~\bibnamefont{Lompe}},
  \bibinfo{author}{\bibfnamefont{T.~B.} \bibnamefont{Ottenstein}},
  \bibinfo{author}{\bibfnamefont{F.}~\bibnamefont{Serwane}},
  \bibinfo{author}{\bibfnamefont{K.}~\bibnamefont{Viering}},
  \bibinfo{author}{\bibfnamefont{A.~N.} \bibnamefont{Wenz}},
  \bibinfo{author}{\bibfnamefont{G.}~\bibnamefont{Z\"urn}}, \bibnamefont{and}
  \bibinfo{author}{\bibfnamefont{S.}~\bibnamefont{Jochim}},
  \bibinfo{journal}{Phys. Rev. Lett.} \textbf{\bibinfo{volume}{105}},
  \bibinfo{pages}{103201} (\bibinfo{year}{2010}),
  \urlprefix\url{http://link.aps.org/doi/10.1103/PhysRevLett.105.103201}.

\bibitem[{\citenamefont{Knoop et~al.}(2009)\citenamefont{Knoop, Ferlaino, Mark,
  Berninger, Sch{\"o}bel, N{\"a}gerl, and Grimm}}]{Natphys5_227}
\bibinfo{author}{\bibfnamefont{S.}~\bibnamefont{Knoop}},
  \bibinfo{author}{\bibfnamefont{F.}~\bibnamefont{Ferlaino}},
  \bibinfo{author}{\bibfnamefont{M.}~\bibnamefont{Mark}},
  \bibinfo{author}{\bibfnamefont{M.}~\bibnamefont{Berninger}},
  \bibinfo{author}{\bibfnamefont{H.}~\bibnamefont{Sch{\"o}bel}},
  \bibinfo{author}{\bibfnamefont{H.-C.} \bibnamefont{N{\"a}gerl}},
  \bibnamefont{and} \bibinfo{author}{\bibfnamefont{R.}~\bibnamefont{Grimm}},
  \bibinfo{journal}{Nature Physics} \textbf{\bibinfo{volume}{5}},
  \bibinfo{pages}{227} (\bibinfo{year}{2009}).

\bibitem[{\citenamefont{Zaccanti et~al.}(2009)\citenamefont{Zaccanti, Deissler,
  D’Errico, Fattori, Jona-Lasinio, M{\"u}ller, Roati, Inguscio, and
  Modugno}}]{Natphys5_586}
\bibinfo{author}{\bibfnamefont{M.}~\bibnamefont{Zaccanti}},
  \bibinfo{author}{\bibfnamefont{B.}~\bibnamefont{Deissler}},
  \bibinfo{author}{\bibfnamefont{C.}~\bibnamefont{D’Errico}},
  \bibinfo{author}{\bibfnamefont{M.}~\bibnamefont{Fattori}},
  \bibinfo{author}{\bibfnamefont{M.}~\bibnamefont{Jona-Lasinio}},
  \bibinfo{author}{\bibfnamefont{S.}~\bibnamefont{M{\"u}ller}},
  \bibinfo{author}{\bibfnamefont{G.}~\bibnamefont{Roati}},
  \bibinfo{author}{\bibfnamefont{M.}~\bibnamefont{Inguscio}}, \bibnamefont{and}
  \bibinfo{author}{\bibfnamefont{G.}~\bibnamefont{Modugno}},
  \bibinfo{journal}{Nature Physics} \textbf{\bibinfo{volume}{5}},
  \bibinfo{pages}{586} (\bibinfo{year}{2009}).

\bibitem[{\citenamefont{Gross et~al.}(2009)\citenamefont{Gross, Shotan,
  Kokkelmans, and Khaykovich}}]{PRL103_Gross}
\bibinfo{author}{\bibfnamefont{N.}~\bibnamefont{Gross}},
  \bibinfo{author}{\bibfnamefont{Z.}~\bibnamefont{Shotan}},
  \bibinfo{author}{\bibfnamefont{S.}~\bibnamefont{Kokkelmans}},
  \bibnamefont{and}
  \bibinfo{author}{\bibfnamefont{L.}~\bibnamefont{Khaykovich}},
  \bibinfo{journal}{Phys. Rev. Lett.} \textbf{\bibinfo{volume}{103}},
  \bibinfo{pages}{163202} (\bibinfo{year}{2009}),
  \urlprefix\url{http://link.aps.org/doi/10.1103/PhysRevLett.103.163202}.

\bibitem[{\citenamefont{Zenesini et~al.}(2014)\citenamefont{Zenesini, Huang,
  Berninger, N\"agerl, Ferlaino, and Grimm}}]{PRA90_Zenesini}
\bibinfo{author}{\bibfnamefont{A.}~\bibnamefont{Zenesini}},
  \bibinfo{author}{\bibfnamefont{B.}~\bibnamefont{Huang}},
  \bibinfo{author}{\bibfnamefont{M.}~\bibnamefont{Berninger}},
  \bibinfo{author}{\bibfnamefont{H.-C.} \bibnamefont{N\"agerl}},
  \bibinfo{author}{\bibfnamefont{F.}~\bibnamefont{Ferlaino}}, \bibnamefont{and}
  \bibinfo{author}{\bibfnamefont{R.}~\bibnamefont{Grimm}},
  \bibinfo{journal}{Phys. Rev. A} \textbf{\bibinfo{volume}{90}},
  \bibinfo{pages}{022704} (\bibinfo{year}{2014}),
  \urlprefix\url{http://link.aps.org/doi/10.1103/PhysRevA.90.022704}.

\bibitem[{\citenamefont{Hu et~al.}(2014)\citenamefont{Hu, Bloom, Jin, and
  Goldwin}}]{PRA90_Hu}
\bibinfo{author}{\bibfnamefont{M.-G.} \bibnamefont{Hu}},
  \bibinfo{author}{\bibfnamefont{R.~S.} \bibnamefont{Bloom}},
  \bibinfo{author}{\bibfnamefont{D.~S.} \bibnamefont{Jin}}, \bibnamefont{and}
  \bibinfo{author}{\bibfnamefont{J.~M.} \bibnamefont{Goldwin}},
  \bibinfo{journal}{Phys. Rev. A} \textbf{\bibinfo{volume}{90}},
  \bibinfo{pages}{013619} (\bibinfo{year}{2014}),
  \urlprefix\url{http://link.aps.org/doi/10.1103/PhysRevA.90.013619}.

\bibitem[{\citenamefont{Dyke et~al.}(2013)\citenamefont{Dyke, Pollack, and
  Hulet}}]{PRA88_Dyke}
\bibinfo{author}{\bibfnamefont{P.}~\bibnamefont{Dyke}},
  \bibinfo{author}{\bibfnamefont{S.~E.} \bibnamefont{Pollack}},
  \bibnamefont{and} \bibinfo{author}{\bibfnamefont{R.~G.} \bibnamefont{Hulet}},
  \bibinfo{journal}{Phys. Rev. A} \textbf{\bibinfo{volume}{88}},
  \bibinfo{pages}{023625} (\bibinfo{year}{2013}),
  \urlprefix\url{http://link.aps.org/doi/10.1103/PhysRevA.88.023625}.

\bibitem[{\citenamefont{Wang and Esry}(2011)}]{NJP_Wang}
\bibinfo{author}{\bibfnamefont{Y.}~\bibnamefont{Wang}} \bibnamefont{and}
  \bibinfo{author}{\bibfnamefont{B.~D.} \bibnamefont{Esry}},
  \bibinfo{journal}{New J. Phys.} \textbf{\bibinfo{volume}{13}},
  \bibinfo{pages}{035025} (\bibinfo{year}{2011}),
  \urlprefix\url{http://stacks.iop.org/1367-2630/13/i=3/a=035025}.

\bibitem[{\citenamefont{Wang et~al.}(2010)\citenamefont{Wang, D'Incao,
  N\"agerl, and Esry}}]{PRL104_Wang}
\bibinfo{author}{\bibfnamefont{Y.}~\bibnamefont{Wang}},
  \bibinfo{author}{\bibfnamefont{J.~P.} \bibnamefont{D'Incao}},
  \bibinfo{author}{\bibfnamefont{H.-C.} \bibnamefont{N\"agerl}},
  \bibnamefont{and} \bibinfo{author}{\bibfnamefont{B.~D.} \bibnamefont{Esry}},
  \bibinfo{journal}{Phys. Rev. Lett.} \textbf{\bibinfo{volume}{104}},
  \bibinfo{pages}{113201} (\bibinfo{year}{2010}),
  \urlprefix\url{http://link.aps.org/doi/10.1103/PhysRevLett.104.113201}.

\bibitem[{\citenamefont{Eismann et~al.}(2016)\citenamefont{Eismann, Khaykovich,
  Laurent, Ferrier-Barbut, Rem, Grier, Delehaye, Chevy, Salomon, Ha
  et~al.}}]{prx-2016-salomon}
\bibinfo{author}{\bibfnamefont{U.}~\bibnamefont{Eismann}},
  \bibinfo{author}{\bibfnamefont{L.}~\bibnamefont{Khaykovich}},
  \bibinfo{author}{\bibfnamefont{S.}~\bibnamefont{Laurent}},
  \bibinfo{author}{\bibfnamefont{I.}~\bibnamefont{Ferrier-Barbut}},
  \bibinfo{author}{\bibfnamefont{B.~S.} \bibnamefont{Rem}},
  \bibinfo{author}{\bibfnamefont{A.~T.} \bibnamefont{Grier}},
  \bibinfo{author}{\bibfnamefont{M.}~\bibnamefont{Delehaye}},
  \bibinfo{author}{\bibfnamefont{F.}~\bibnamefont{Chevy}},
  \bibinfo{author}{\bibfnamefont{C.}~\bibnamefont{Salomon}},
  \bibinfo{author}{\bibfnamefont{L.-C.} \bibnamefont{Ha}},
  \bibnamefont{et~al.}, \bibinfo{journal}{Phys. Rev. X}
  \textbf{\bibinfo{volume}{6}}, \bibinfo{pages}{021025} (\bibinfo{year}{2016}),
  \urlprefix\url{http://link.aps.org/doi/10.1103/PhysRevX.6.021025}.

\bibitem[{\citenamefont{Rem et~al.}(2013)\citenamefont{Rem, Grier,
  Ferrier-Barbut, Eismann, Langen, Navon, Khaykovich, Werner, Petrov, Chevy
  et~al.}}]{prl-2013-salomon}
\bibinfo{author}{\bibfnamefont{B.~S.} \bibnamefont{Rem}},
  \bibinfo{author}{\bibfnamefont{A.~T.} \bibnamefont{Grier}},
  \bibinfo{author}{\bibfnamefont{I.}~\bibnamefont{Ferrier-Barbut}},
  \bibinfo{author}{\bibfnamefont{U.}~\bibnamefont{Eismann}},
  \bibinfo{author}{\bibfnamefont{T.}~\bibnamefont{Langen}},
  \bibinfo{author}{\bibfnamefont{N.}~\bibnamefont{Navon}},
  \bibinfo{author}{\bibfnamefont{L.}~\bibnamefont{Khaykovich}},
  \bibinfo{author}{\bibfnamefont{F.}~\bibnamefont{Werner}},
  \bibinfo{author}{\bibfnamefont{D.~S.} \bibnamefont{Petrov}},
  \bibinfo{author}{\bibfnamefont{F.}~\bibnamefont{Chevy}},
  \bibnamefont{et~al.}, \bibinfo{journal}{Phys. Rev. Lett.}
  \textbf{\bibinfo{volume}{110}}, \bibinfo{pages}{163202}
  (\bibinfo{year}{2013}),
  \urlprefix\url{http://link.aps.org/doi/10.1103/PhysRevLett.110.163202}.

\bibitem[{\citenamefont{S\o{}rensen et~al.}(2013)\citenamefont{S\o{}rensen,
  Fedorov, Jensen, and Zinner}}]{pra-2013-zinner}
\bibinfo{author}{\bibfnamefont{P.~K.} \bibnamefont{S\o{}rensen}},
  \bibinfo{author}{\bibfnamefont{D.~V.} \bibnamefont{Fedorov}},
  \bibinfo{author}{\bibfnamefont{A.~S.} \bibnamefont{Jensen}},
  \bibnamefont{and} \bibinfo{author}{\bibfnamefont{N.~T.}
  \bibnamefont{Zinner}}, \bibinfo{journal}{Phys. Rev. A}
  \textbf{\bibinfo{volume}{88}}, \bibinfo{pages}{042518}
  (\bibinfo{year}{2013}),
  \urlprefix\url{http://link.aps.org/doi/10.1103/PhysRevA.88.042518}.

\bibitem[{\citenamefont{Huang et~al.}(2015)\citenamefont{Huang, Sidorenkov, and
  Grimm}}]{pra-2015-grimm}
\bibinfo{author}{\bibfnamefont{B.}~\bibnamefont{Huang}},
  \bibinfo{author}{\bibfnamefont{L.~A.} \bibnamefont{Sidorenkov}},
  \bibnamefont{and} \bibinfo{author}{\bibfnamefont{R.}~\bibnamefont{Grimm}},
  \bibinfo{journal}{Phys. Rev. A} \textbf{\bibinfo{volume}{91}},
  \bibinfo{pages}{063622} (\bibinfo{year}{2015}),
  \urlprefix\url{http://link.aps.org/doi/10.1103/PhysRevA.91.063622}.

\bibitem[{\citenamefont{Wang et~al.}(2013)\citenamefont{Wang, D'Incao, and
  Esry}}]{long_review}
\bibinfo{author}{\bibfnamefont{Y.}~\bibnamefont{Wang}},
  \bibinfo{author}{\bibfnamefont{J.~P.} \bibnamefont{D'Incao}},
  \bibnamefont{and} \bibinfo{author}{\bibfnamefont{B.~D.} \bibnamefont{Esry}}
  (\bibinfo{publisher}{Academic Press}, \bibinfo{year}{2013}),
  vol.~\bibinfo{volume}{62} of \emph{\bibinfo{series}{Adv. At. Mol. Opt.
  Phys.}}, chap.~\bibinfo{chapter}{1}, pp. \bibinfo{pages}{1 -- 115},
  \urlprefix\url{http://www.sciencedirect.com/science/article/pii/B9780124080904000013}.

\bibitem[{\citenamefont{Taylor}(2012)}]{Taylor}
\bibinfo{author}{\bibfnamefont{J.~R.} \bibnamefont{Taylor}},
  \emph{\bibinfo{title}{Scattering Theory}} (\bibinfo{publisher}{Dover
  Publications}, \bibinfo{year}{2012}), ISBN \bibinfo{isbn}{9780486142074}.

\bibitem[{\citenamefont{Simbotin and C\^ot\'e}(2015)}]{Jost_paper}
\bibinfo{author}{\bibfnamefont{I.}~\bibnamefont{Simbotin}} \bibnamefont{and}
  \bibinfo{author}{\bibfnamefont{R.}~\bibnamefont{C\^ot\'e}},
  \bibinfo{journal}{Chem. Phys.} \textbf{\bibinfo{volume}{462}},
  \bibinfo{pages}{79 } (\bibinfo{year}{2015}), ISSN \bibinfo{issn}{0301-0104},
  \urlprefix\url{http://www.sciencedirect.com/science/article/pii/S0301010415001834}.

\bibitem[{\citenamefont{D'Incao and Esry}(2005)}]{PRL94_Incao}
\bibinfo{author}{\bibfnamefont{J.~P.} \bibnamefont{D'Incao}} \bibnamefont{and}
  \bibinfo{author}{\bibfnamefont{B.~D.} \bibnamefont{Esry}},
  \bibinfo{journal}{Phys. Rev. Lett.} \textbf{\bibinfo{volume}{94}},
  \bibinfo{pages}{213201} (\bibinfo{year}{2005}),
  \urlprefix\url{http://link.aps.org/doi/10.1103/PhysRevLett.94.213201}.

\bibitem[{\citenamefont{Olver and Maximon}(2010)}]{nist_bessel_hand_book}
\bibinfo{author}{\bibfnamefont{F.~W.~J.} \bibnamefont{Olver}} \bibnamefont{and}
  \bibinfo{author}{\bibfnamefont{L.~C.} \bibnamefont{Maximon}},
  \emph{\bibinfo{title}{NIST Handbook of Mathematical Functions}}
  (\bibinfo{publisher}{Cambridge University Press}, \bibinfo{year}{2010}),
  chap.~\bibinfo{chapter}{10}, ISBN \bibinfo{isbn}{9780521192255},
  \urlprefix\url{https://books.google.com/books?id=3I15Ph1Qf38C}.

\end{thebibliography}

\end{document}